\documentclass[]{article}
\usepackage{arxiv}

\usepackage{graphicx} 

\usepackage{hyperref}
\usepackage{caption}
\usepackage{subcaption}
\usepackage{atbegshi}
\AtBeginDocument{\AtBeginShipoutNext{\AtBeginShipoutDiscard}}

\usepackage{amsmath,amssymb}
\usepackage{nicefrac}
\usepackage[capitalise]{cleveref}
\usepackage{bm,upgreek}
\usepackage{adjustbox}
\usepackage{booktabs}
\usepackage{appendix}
\Crefname{appsec}{Appendix}{appendices}
\numberwithin{equation}{section}
\usepackage{lineno, blindtext}
\usepackage[font=small,labelfont=bf]{caption}
\usepackage{color}
\usepackage{multirow}
\usepackage{colortbl}
\usepackage[table]{xcolor}
\usepackage{float}
\usepackage{array} 
\usepackage{ragged2e} 
\usepackage{tabularx} 
\usepackage{booktabs}
\usepackage{xcolor}

\newcommand{\ppap}[1]{{\color{black}{#1}}}
\newcommand{\shorttitle}{Loachamin et al. (2024)}

\usepackage{lineno}

\title{Discovering Deposition Process Regimes:\\ Leveraging Unsupervised Learning for Process Insights, Surrogate Modeling, and Sensitivity Analysis}

\author{\href{https://orcid.org/0009-0002-4531-7960}{\includegraphics[scale=0.06]{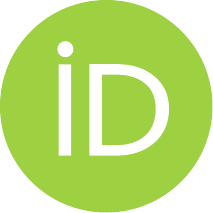}\hspace{1mm}Geremy Loachamin Suntaxi\thanks{Author also affiliated with the School of Chemical Engineering, National Technical University of Athens, Zographos Campus, 15780, Attiki, Greece}}\\
	Faculty of Science, Technology and Medicine\\
	University of Luxembourg\\
	Esch-sur-Alzette, L-4364, Luxembourg\\
    \texttt{geremy.loachamin@uni.lu}
    \And
  \href{https://orcid.org/0000-0002-3136-1402}{\includegraphics[scale=0.06]{orcid.pdf}\hspace{1mm}Paris Papavasileiou\thanks{Author also affiliated with the School of Chemical Engineering, National Technical University of Athens, Zographos Campus, 15780, Attiki, Greece}}\\
	Faculty of Science, Technology and Medicine\\
	University of Luxembourg\\
	Esch-sur-Alzette, L-4364, Luxembourg\\
    \texttt{paris.papavasileiou@uni.lu}
    \And
    \href{https://orcid.org/0000-0002-5229-4157}{\includegraphics[scale=0.06]{orcid.pdf}\hspace{1mm}Eleni D.~Koronaki} \\
	Faculty of Science, Technology and Medicine\\
	University of Luxembourg\\
	Esch-sur-Alzette, L-4364, Luxembourg\\
	\texttt{eleni.koronaki@uni.lu}
	\And
	\href{https://orcid.org/0000-0003-2272-2584}
    {\includegraphics[scale=0.06]{orcid.pdf}\hspace{1mm}Dimitrios G.~Giovanis} \\
	Department of Civil \& Systems Engineering \\
    Johns Hopkins University\\
    Baltimore, MD 21218, USA\\
	\texttt{dgiovan1@jhu.edu}\\
     \And
    \href{https://orcid.org/0000-0002-0558-9972}{\includegraphics[scale=0.06]{orcid.pdf}\hspace{1mm}Georgios Gakis} \\
	School of Chemical Engineering\\
    National Technical University of Athens\\
    Zographos Campus, 15780, Attiki, Greece\\
	\texttt{gakisg@chemeng.ntua.gr}
    \And
    Ioannis G.~Aviziotis \\
	School of Chemical Engineering\\
    National Technical University of Athens\\
    Zographos Campus, 15780, Attiki, Greece\\
	\texttt{javiziot@chemeng.ntua.gr}
    \And
	Martin Kathrein \\
	CERATIZIT Luxembourg S.à r.l.\\
    Mamer, L-8201, Luxembourg\\
	\texttt{Martin.Kathrein@ceratizit.com} \\
    \And
	Gabriele Pozzetti \\
	CERATIZIT Luxembourg S.à r.l.\\
    Mamer, L-8201, Luxembourg\\
	\texttt{Gabriele.Pozzetti@plansee-group.com}\\
    \And
	Christoph Czettl \\
	CERATIZIT Austria GmbH\\
    Reutte, A-6600, Austria\\
	\texttt{Christoph.Czettl@ceratizit.com} \\ 
    \And
	\href{https://orcid.org/0000-0001-7622-2193}{\includegraphics[scale=0.06]{orcid.pdf}\hspace{1mm}St\'{e}phane P.A.~Bordas}\thanks{corresponding author: stephane.bordas@uni.lu} \\
	Faculty of Science, Technology and Medicine\\
	University of Luxembourg\\
	Esch-sur-Alzette, L-4364 \\
	\texttt{stephane.bordas@uni.lu}  \\ 
    \And
    \href{https://orcid.org/0000-0001-6651-7318}{\includegraphics[scale=0.06]{orcid.pdf}\hspace{1mm}Andreas G.~Boudouvis} \\
	School of Chemical Engineering\\
    National Technical University of Athens\\
    Zographos Campus, 15780, Attiki, Greece\\
	\texttt{boudouvi@chemeng.ntua.gr}
}

\begin{document}

\maketitle
\begin{abstract}
This work introduces a comprehensive approach utilizing data-driven methods to elucidate the deposition process regimes in Chemical Vapor Deposition (CVD) reactors and the interplay of physical mechanism that dominate in each one of them. Through this work, we address three key objectives. Firstly, our methodology relies on process outcomes, derived by a detailed CFD model, to identify clusters of "outcomes" corresponding to distinct process regimes, wherein the relative influence of input variables undergoes notable shifts. This phenomenon is experimentally validated through Arrhenius plot analysis, affirming the efficacy of our approach. Secondly, we demonstrate the development of an efficient surrogate model, based on Polynomial Chaos Expansion (PCE), that maintains accuracy, facilitating streamlined computational analyses. Finally, as a result of PCE,   sensitivity analysis is made possible by means of Sobol' indices, that quantify the impact of process inputs across identified regimes. 

The insights gained from our analysis contribute to the formulation of hypotheses regarding phenomena occurring beyond the transition regime. Notably, the significance of temperature even in the diffusion-limited regime, as evidenced by the Arrhenius plot, suggests activation of gas phase reactions at elevated temperatures. Importantly, our proposed methods yield insights that align with experimental observations and theoretical principles, aiding decision-making in process design and optimization. By circumventing the need for costly and time-consuming experiments, our approach offers a pragmatic pathway towards enhanced process efficiency. Moreover, this study underscores the potential of data-driven computational methods for innovating reactor design paradigms.
\end{abstract}

\maketitle
\section{Introduction}\label{Sec:Intro}
Vapor deposition processes have been extensively used for the production of a wide range of materials in the form of coatings and thin films \cite{choy2003chemical,george2010atomic,helmersson2006ionized}. During the last decades, the emergence of nanotechnology has extended the use of such techniques for the production of semiconductors \cite{4melloch1997compound}, optoelectronics \cite{5gkinis2017effects}, high-k dielectrics \cite{6gakis2019investigation}, functional powder coatings \cite{7vahlas2006principles}, 2D materials \cite{8suk2011transfer}, as well as nanostructures, such as carbon nanotubes \cite{9gakis2022unraveling}. Among the various deposition techniques, Chemical Vapor Deposition (CVD) is often preferred due to its versatility, high throughput, relatively low cost, and better control over the deposition uniformity and conformality \cite{choy2003chemical,10psarellis2018investigation}.

Besides the wide use and advantages of CVD, the process design and scale-up for industrial applications is often challenging, owing to the complex nature of the different mechanisms and phenomena that take place during the process \cite{choy2003chemical,koronaki2016efficient}. A typical CVD process incorporates precursor gas flows and mass transfer, precursor reactions and decomposition in the gas phase, as well as surface phenomena, such as adsorption, desorption and surface reactions \cite{11kleijn2007multi}. The interplay between these phenomena may affect the rate, uniformity and conformality of the deposition \cite{5gkinis2017effects,12gakis2019detailed,gkinis2019skouteris}, as well as the decomposition of the deposited product, thus having an adverse effect on its properties \cite{9gakis2022unraveling}. As the contributions of these mechanisms and phenomena depend on the process conditions, a specified range of conditions is often provided for the efficient deposition of films with desired properties.   

In CVD processes, the effect of process conditions is typically studied using the Arrhenius plot, showing the deposition rate as a function of temperature. Usually, two regimes are identified \cite{choy2003chemical}: (i) the reaction-limited regime, where the reactions are slow and hence the rate-limiting step, and (ii) the diffusion-limited regime, where the mass transfer is slower than the reactions and therefore is the rate-limiting step. In the reaction-limited regime, the deposition rate increases with temperature, while in the diffusion-limited regime, which occurs above a certain critical temperature, the deposition rate can be unaffected or slightly decrease with a further temperature increase. 

Most of the recipes used in the industrial application of such processes are based on the selection of process conditions that lead to the required regime. However, the regime definition from the Arrhenius plot is not always straightforward. Several studies have shown that other mechanisms may limit the deposition at higher temperatures besides the precursor mass transfer, such as precursor depletion \cite{10psarellis2018investigation}, pre-decomposition in the gas phase \cite{13fritzsche2021atmospheric}, desorption \cite{12gakis2019detailed,14aviziotis2017combined}, or catalyst deactivation \cite{9gakis2022unraveling,15gakis2023multi}, among others. Often, the number of regimes is larger, with transition regimes also identified \cite{13fritzsche2021atmospheric,14aviziotis2017combined}. Therefore, the choice of the appropriate process conditions requires the identification and mapping of these regimes on the process parameter space. As these complex regimes are an outcome of the interplay between various mechanisms, this demands a thorough study of the effect of process parameters on these mechanisms, in order to obtain optimal sets of parameters. 

The rigorous investigation of such a complex process is an experimentally challenging task. Therefore, computational modeling of CVD processes has been employed to provide an insight of the different mechanisms that take place \cite{15gakis2023multi,16gakis2015numerical,17papavasileiou2022efficient}. Using combined experimental and computational approaches, it is possible not only to identify the mechanisms and phenomena that determine the deposition process \cite{10psarellis2018investigation, 12gakis2019detailed}, but also to study the effect of process parameters on such mechanisms \cite{15gakis2023multi}, in order to provide a process parameter space for optimal process design. Nevertheless, although combined experimental and computational mechanistic approaches can provide a complete insight and understanding of CVD processes, they remain time-consuming, and demanding in terms of experimental and computational cost, as well as in terms of scientific information required as input, such as chemical reactions, their rate constants, and activation energies. Therefore, the efficient identification of the process conditions that correspond to each regime in the Arrhenius plot remains an open problem. 

Data-driven approaches have been introduced, to provide predictive and robust models for CVD processes \cite{18papavasileiou2023equation,19koronaki2020data}. Such models may use either experimental or computational data \cite{20spencer2021investigation} as input in order to predict process outputs or parameters \cite{21koronaki2023partial}. The data-driven models have lower computational cost than the mechanistic models \cite{18papavasileiou2023equation,martin2023physics}, while they do not require scientific information (such as reaction parameters) as input to complement the data. Along these lines, uncertainty quantification and sensitivity analysis of complex systems in the context of chemical processes, as well as data-driven surrogate models have become outstanding methods for emulating such systems with high accuracy and low computational cost, and for providing physical insight \cite{SURRO18, SURRO19}. Although these methods have been extensively studied, when faced with applications related to data-driven surrogate modeling for industrial-scale processes, some important obstacles and challenges remain to be addressed.

\begin{enumerate}
	\item The amount and quality of data collected (either by experiments or simulations). In industrial-scale experiments, these (high-fidelity) data can be extremely costly to obtain, leading to a reduced amount of information. Moreover, as far as computational models are concerned, a factor to be considered is the ratio between the number of variables and the number of simulations necessary to train a data-driven model, which generally implies a high computational cost.

	\item Concerning the Arrhenius plots, there is lack of analytical criteria to distinguish the different regimes especially in industrial-scale processes. Furthermore, the imbalance of dataset size in the various regimes introduces further obstacles.
 
	\item Proper quantification of the influence of input variability on the output responses of models that mimic complex industrial-scale processes to determine their importance in the process and to provide them with a physical sense that can serve as criteria for system optimization.
\end{enumerate}	

The ambition of this work is to propose a data-driven workflow based on unsupervised methods, in order to deliver predictions of the relative effect of the physical phenomena on the final process outcome. The goal is to avoid the time and cost involved in the line of experiments typically required for determining the Arrhenius plot of a deposition process. Moreover, an accurate and efficient surrogate modeling approach is presented, which further allows to perform sensitivity analysis of the input parameters. Here the validation of our model via comparison to experimental data, confirms the fundamental theory of Fe deposition. This sets stage for the ultimate goal, which is to develop a computational tool that enables efficient design of novel processes or the optimization of existing industrial set-ups, where data production are already available.

\section{Methods}
\subsection{Process description and experimental set up} \label{Sec:Process}

Fe deposition was conducted in a vertical, cylindrical, stagnant flow, cold wall CVD reactor which has been previously described in detail \cite{14aviziotis2017combined, xenidou2010reaction}. 
\begin{figure}[ht!]
 	\centering
	\includegraphics[width=0.6 \linewidth]{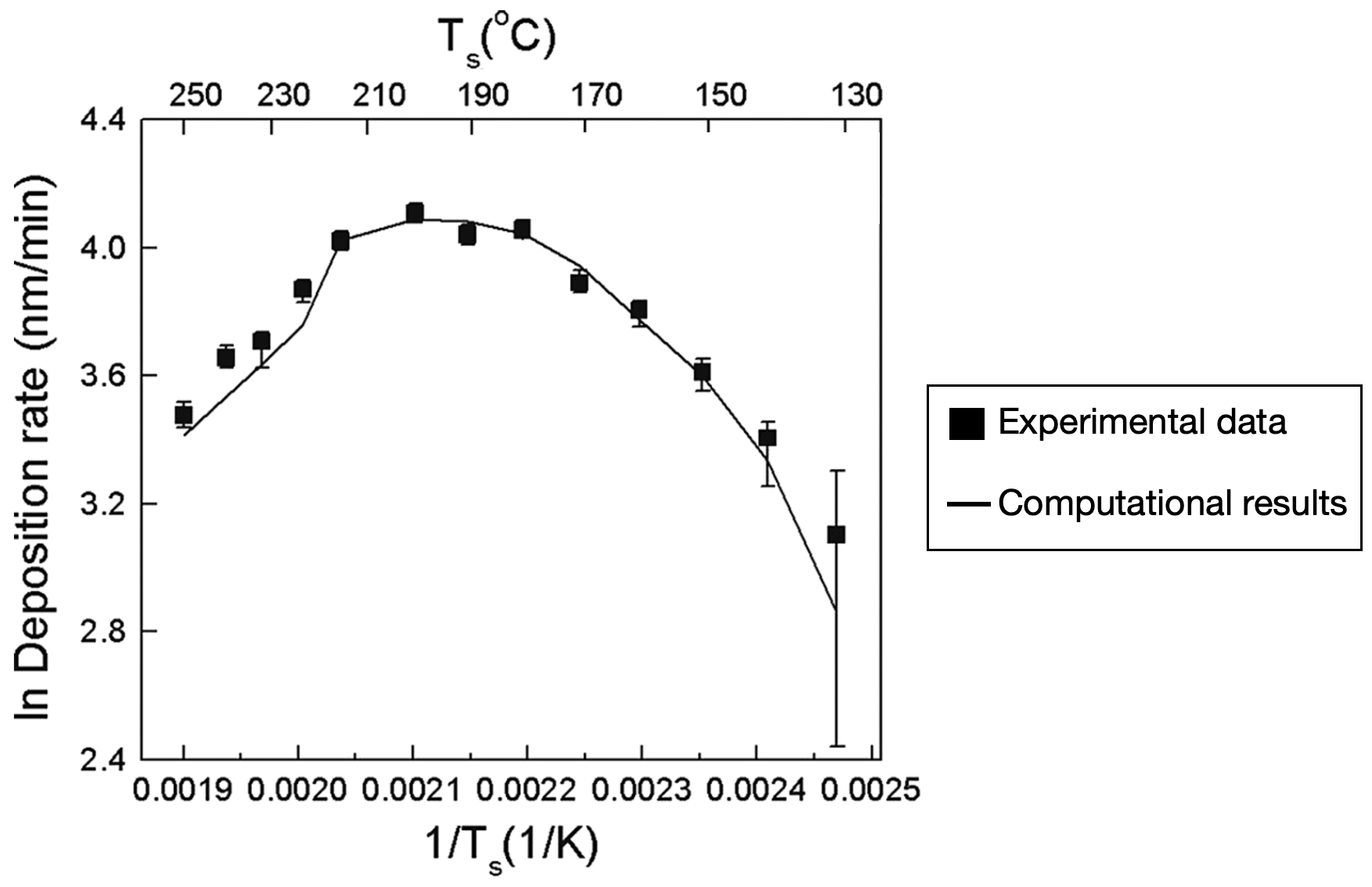}
	\caption{The Arrhenius plot of a CVD reactor for the Fe from Fe(CO)$_5$ and a fixed pressure $P = 10$ Torr \cite{14aviziotis2017combined}. Experimental measurements (squares) and computational results (line) are shown.}\label{Fig:Arrh} 
\end{figure}

Independent experiments were conducted: (i) at 13 different surface temperature values, $T_s$, in the range $130-250$ °C for a fixed pressure of $P_\text{reactor} = 10$ Torr, and (ii) at four different $P_\text{reactor}$, in the range $10-40$ Torr, for a fixed $T_s = 180$ °C. The deposition duration for all experiments was 1 h. 10 x 10 x 1 $mm^3$ silicon coupons are cut from 4 in Si(100) wafers (Sil’tronix). They are cleaned in an ultrasonic bath with acetone and ethanol, they are dried under argon flux and baked in a furnace at $60$ °C for 30 min to cast off humidity. The deposition rate is evaluated directly by weight difference of the substrates before and after deposition, using a microbalance (Sartorius) with $\pm 10 \mu g$ accuracy. Three independent weight measurements are carried out for each substrate before and after the experiment and an average value is calculated. The Fe precursor, Fe(CO)$_5$, is supplied by Sigma-Aldrich and Fischer Scientific and is used as received. Pure nitrogen ($99,998\%$, Air Products) is fed through computer-driven mass flow controllers. During experiments, the lines and walls of the reactor are maintained at room temperature (regulated at 25 °C), while the $N_2$ dilution gas flow and the $N_2$ carrier gas flow through the precursor equal 302 and 3 standard cubic centimeters per minute (sccm), respectively. 

\subsection{CFD model and Data collection}\label{Sec:CFD}
\ppap{A CFD model was developed based on the work of Aviziotis et al. \cite{aviziotisChemicalVaporDeposition2016, 14aviziotis2017combined} for the CVD of Fe from Fe(CO)$_5$. The original CFD model \cite{14aviziotis2017combined} is 3D, its equations are discretized using the finite volume method and solved using Fluent CFD code. In this work, we construct a 2D$-$axisymmetric CFD model, the equations are then discretized using 13987 finite elements and solved using COMSOL Multiphysics\textsuperscript{\tiny{\textregistered}}.

The governing equations include the conservation of mass, momentum, and energy, along with the conservation equations of the chemical species that partake in the chemical reactions. For more details, the interested reader is referred to \cite{aviziotisChemicalVaporDeposition2016}.

A constant mass inflow rate of $5.9177 \cdot 10^{-6}$ kg/s is imposed at the inlet of the reactor. The inlet stream consists solely of $N_2$ and the precursor (Fe(CO)$_5$). At the outlet boundary, the pressure is set equal to the operating pressure of the reactor. A no-slip condition is set for all of the inside walls of the reactor. The flow is considered laminar.

A constant temperature is set for the substrate surface ($T_s$), the side walls of the reactor ($T_w$ = 25 °C), and the gas reactant lines ($T_l$ = 25 °C). The ideal gas assumption is made for calculating the density of the gas mixture. The species’ heat capacity, molar enthalpy, and molar entropy are calculated using the relevant CHEMKIN coefficients and NASA polynomials. Furthermore, a Stefan–Maxwell diffusion model is used to compute the diffusion fluxes.

The gas phase reactions are summarized in Table \ref{Tab:gas-phase-reactions}. They consist of four decomposition (forward) reactions ($G1-G4$) and three recombination (reverse) reactions ($G1'-G3'$). The reaction rates for the forward reactions ($R_{G_i}$) can be calculated as such:
\begin{equation}
    R_{G_i}=k_{0,G_i}\ \text{exp}\left(\nicefrac{-E_{a,G_i}}{RT}\right) C_{\text{Fe(CO)}_{i}},
\end{equation}
and for the reverse reactions ($R_{G'_i}$):
\begin{equation}
    R_{G'_i}=k_{0,G'_i}\ \text{exp}\left(\nicefrac{-E_{a,G'_i}}{RT}\right) C_{\text{Fe(CO)}_{i}}C_{\text{CO}}.
\end{equation}

\begin{table}[!ht]
\centering
\caption{\ppap{Gas phase reactions along with their pre-exponential factors and activation energies. The forward reactions are denoted as $G1-G4$ and the reverse reactions as $G1'-G3'$.}}
\begin{adjustbox}{width=1\textwidth}
\begin{tabular}{l l c c}
\hline
ID  & Reaction               & Pre-exponential factor ($k_0$)   & Activation energy ($E_a$) [kJ mol\textsuperscript{-1}] \\
\hline
G1  & Fe(CO)\textsubscript{5} → Fe(CO)\textsubscript{4} + CO & $9.65 \cdot 10^{12}$ $\mathrm{s^{-1}}$           & 136.7       \\
G2  & Fe(CO)\textsubscript{4} → Fe(CO)\textsubscript{3} + CO & $8.96 \cdot 10^{12}$ $\mathrm{s^{-1}}$           & 79.9        \\
G3  & Fe(CO)\textsubscript{3} → Fe(CO)\textsubscript{2} + CO & $1.25 \cdot 10^{11}$ $\mathrm{s^{-1}}$           & 97.5        \\
G4  & Fe(CO)\textsubscript{2} → FeCO + CO                    & $3.96 \cdot 10^{11}$ $\mathrm{s^{-1}}$           & 139.1       \\
$G1'$ & Fe(CO)\textsubscript{4} + CO → Fe(CO)\textsubscript{5} & $3.5 \cdot 10^{7}$ $\mathrm{m^3\,  kmol^{-1}\,  s^{-1}}$ & 10.5\\
$G2'$ & Fe(CO)\textsubscript{3} + CO → Fe(CO)\textsubscript{4} & $1.3 \cdot 10^{10}$ $\mathrm{m^3\,  kmol^{-1}\,  s^{-1}}$ & 9.5\\
$G3'$ & Fe(CO)\textsubscript{2} + CO → Fe(CO)\textsubscript{3} & $1.8 \cdot 10^{10}$ $\mathrm{m^3\,  kmol^{-1}\,  s^{-1}}$ & 9.5\\
\hline
\end{tabular}
\end{adjustbox}
\label{Tab:gas-phase-reactions}
\end{table}

The surface reactions that ultimately lead to the deposition of Fe are summarized in Table \ref{Tab:surface-reactions}. For the rates of these reactions ($R_{S, i}$), a Langmuir-Hinshelwood type kinetic expression is implemented, given by:
\begin{equation}
    R_{S_i}=\nicefrac{\displaystyle k_{0,S_i}\ \text{exp}\left(\nicefrac{-E_{a,S_i}}{RT_s}\right) C_{\text{Fe(CO)}_{i}}\ }{\ \displaystyle (1+k_{CO}\ \text{exp}\left(\nicefrac{-E_{a,CO}}{RT_s}\right)P_{CO})}.
\end{equation}

\begin{table}[!ht]
\centering
\caption{\ppap{Surface reactions along with their pre-exponential factors and activation energies.}}
\begin{adjustbox}{width=1\textwidth}
\begin{tabular}{l l c c}
\hline
ID  & Reaction               & Pre-exponential factor ($k_0$)   & Activation energy ($E_a$) [kJ mol\textsuperscript{-1}] \\
\hline
S1  & Fe(CO)\textsubscript{5} → Fe\textsubscript{(s)} + 5CO\textsubscript{(g)} & $6.00 \cdot 10^{2}$ $\mathrm{m\, s^{-1}}$           & 27.9      \\
S2  & Fe(CO)\textsubscript{3} → Fe\textsubscript{(s)} + 3CO\textsubscript{(g)} & $5.30 \cdot 10^{7}$ $\mathrm{m\, s^{-1}}$            & 75.3        \\
S3  & Fe(CO) → Fe\textsubscript{(s)} + CO\textsubscript{(g)} & $3.70 \cdot 10^{10}$ $\mathrm{m\, s^{-1}}$            & 19.3        \\
S4  & CO adsorption & $1.90 \cdot 10^{11}$ $\mathrm{Torr^{-1}}$ & 89.9 \\
\hline
\end{tabular}
\end{adjustbox}
\label{Tab:surface-reactions}
\end{table}


The kinetic parameters of the model are fitted to the experimental data  \cite{aviziotisChemicalVaporDeposition2016} for temperatures between $130-250$ °C and an operating pressure of $P=10$ Torr.

For pressures higher than 10 Torr, the pre-exponential factor of S1 is adjusted:
\begin{equation}
    k_{{0,S_1},adapted}= \nicefrac{\displaystyle 600 [m/s] \ 20[Torr^2]}{\displaystyle P^2}.
\end{equation}

Following model validation, data are collected for two parameters of interest:
\begin{enumerate}
    \item The temperature of the deposition surface ($T_s$).
    \item The operating pressure of the reactor ($P$).
\end{enumerate}

Computational experiments are conducted for $T_s$ varying from $130-250\ \text{°C}$ and $P$ at values of 10, 20, 30 and 40 Torr. The design of experiments (DoE) is summarized in Table \ref{Tab:DoE} below:

\begin{table}[!ht]
\centering
\caption{\ppap{Range of parameters for data-collection purposes.}}
    \begin{tabular}{l r l r l}
    \hline
    Parameter  & Range     & units          & \# of values & sampling \\
    \hline
    $T_s$ & [130, 250] & \textdegree C & 601 & uniform distribution\\
    $P$ & [10, 40] & Torr & 4 & uniform distribution\\
\hline
\end{tabular}
\label{Tab:DoE}
\end{table}

This DoE leads to a total of 2404 computational experiments.} 

\subsection{Data-driven workflow}\label{Sec:Workflow}
\begin{figure}[ht!]
	\centering
	\includegraphics[width=1 \linewidth]{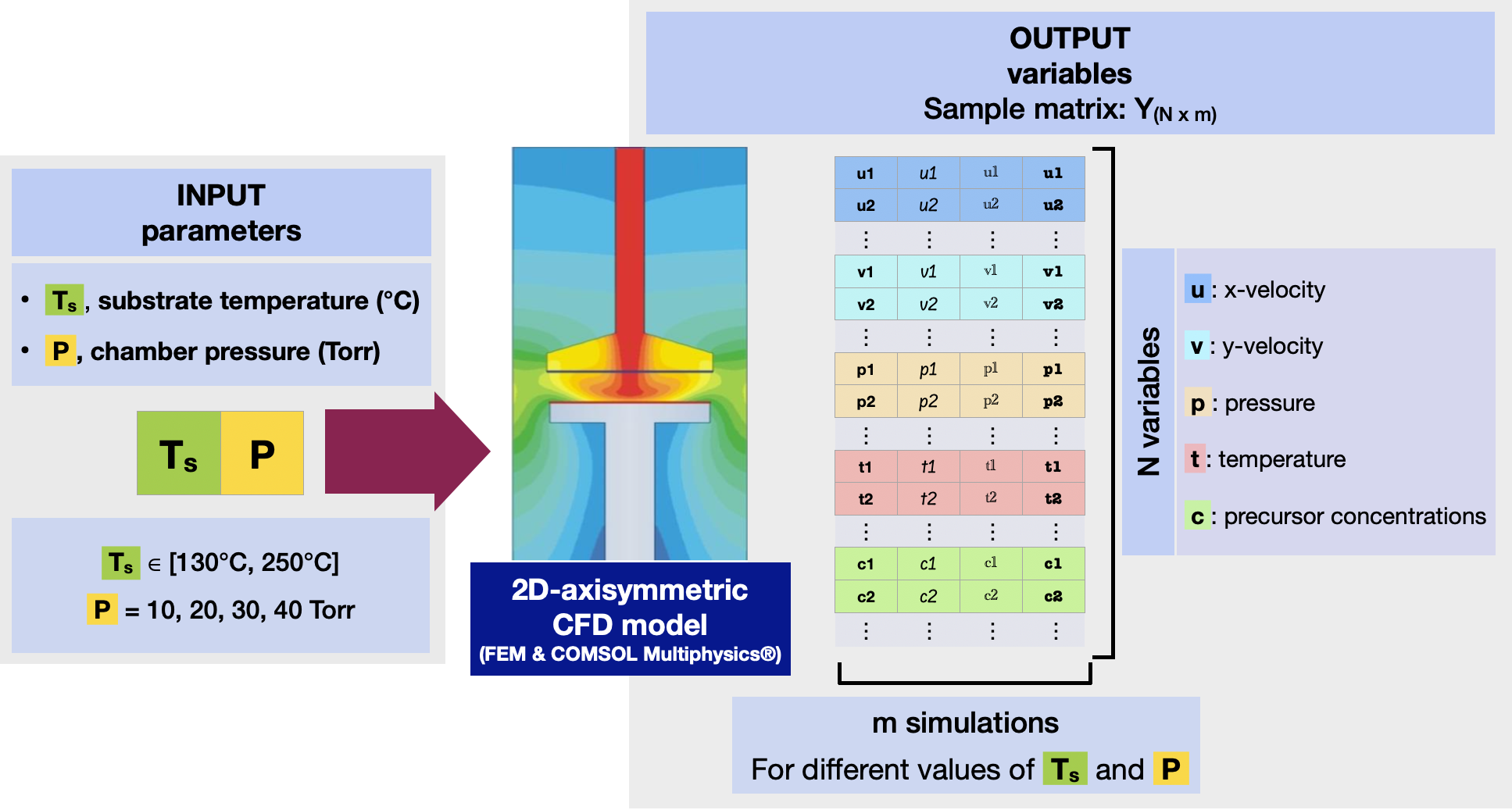}
	\caption{The dataset comes from a CFD model for the Fe CVD reactor. The input parameters are the substrate temperature $T_s$ and the chamber pressure $P$.
	This computational model simulates seven variables at each discretization point, including the distributions of the x-y velocities $u$ and $v$, pressure $p$, temperature $t$, and three different precursor concentrations presented in \cref{Tab:surface-reactions}.}\label{Fig:Datac}
\end{figure}

 The data-collection process involves collecting data from the CFD model based on the parameter values reported in Table \ref{Tab:DoE}. At each combination of parameters\footnote{the set of parameter combinations constitute the experimental design, denoted by $\bm{\mathcal{X}}$.}, the values for the solved variables at each discretization point are arranged in columns, as shown schematically in Fig. \ref{Fig:Datac}. The proposed data-driven workflow is  detailed in Section \ref{Sec:scheme}. But first, we provide a brief overview of the methods applied in this work.
		
    \subsubsection{Dimensionality reduction using Principal component analysis}\label{subsec:PCA}

\begin{figure}[!ht]
		\centering
		\includegraphics[width=1 \linewidth]{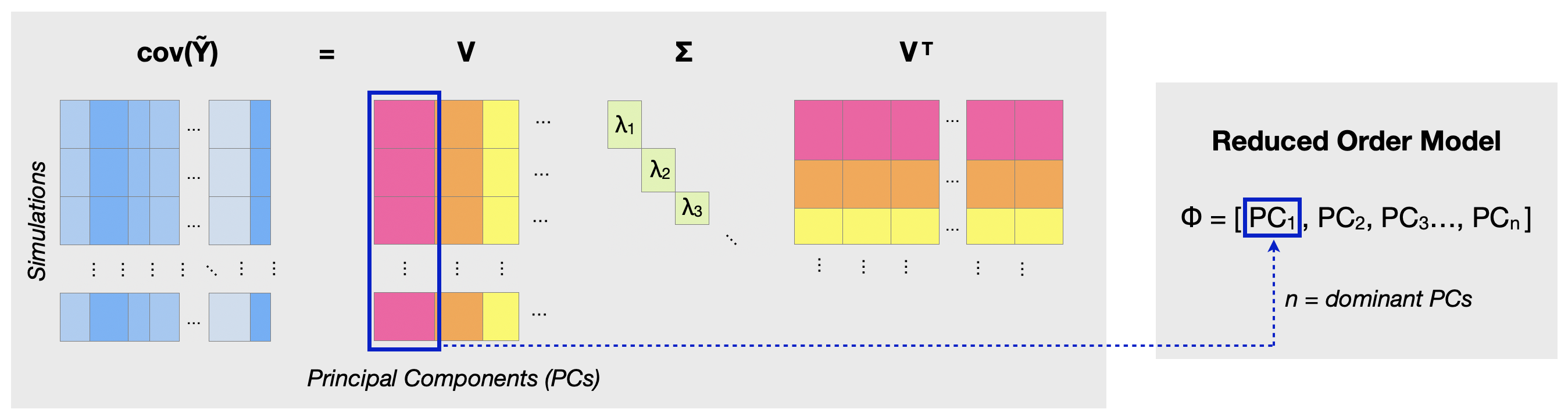}
		\caption{PCA decomposes the covariance matrix of $Y$ (cov($\tilde{Y}$)) into $n$ dominant principal components (PCs): ($PC_1, PC_2, \cdots, PC_n$), from which we can obtain the reduced space $\Upphi$.}\label{Fig:PCA}
	\end{figure}

    Principal component analysis (PCA) is a method widely used for reduction of the dimension of a scaled dataset $\tilde{\textbf{Y}}$ (containing numerical simulations or experimental measurements) \cite{HOTELLING1933, MACKIEWICZ1993}. Essentially, PCA seeks to find a suitable low-dimensional space, based on the \textit{singular value decomposition} (SVD) of the covariance matrix of $\tilde{\textbf{Y}}$, $\text{cov}(\tilde{\textbf{Y}})$ (see \cref{Fig:PCA}), to represent the original high-dimensional data while capturing its total variability. PCA has been successfully applied for dimensionality reduction in the context of CVD processes \cite{ISAAC14, HUANG2019, KORONAKI2019}. The key-ingredients of PCA are  presented in detail in \cref{AppendixA}. 

    \subsubsection{Unsupervised learning using hierarchical (agglomerative) clustering}\label{subsec:Cluster}
	
\begin{figure}[H]
			\centering
			\includegraphics[width=0.4\linewidth]{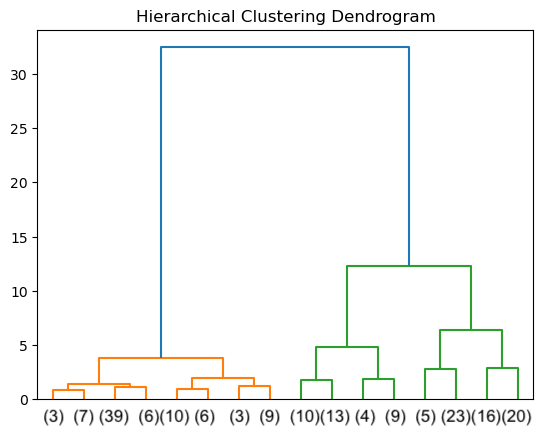}
			\caption{HC gradually merges smaller clusters to form a nested tree structure (dendrogram) where each branch corresponds to a group (or cluster). The three largest clusters are colored, while the number of elements in each cluster is shown in parentheses at the bottom.}\label{fig-clus}
		\end{figure}

	Hierarchical clustering (HC) is an unsupervised machine learning method used to group data based on the similarities between elements in an agglomerative manner \cite{GORDON1987}. The algorithm creates a hierarchical nested tree, known as a dendrogram, from this grouping process (see Fig. \ref{fig-clus}). At the bottom of this tree, we have individual clusters: one cluster for each dataset element. Then, we merge pairs of clusters and order them hierarchically until we reach the top of the tree, where a single cluster contains all the elements. A cut-off of the dendrogram is then chosen at a certain distance level to form several homogeneous groups with common characteristics.
  
	 To determine the distance between groups, we can use several similarity measures depending on the elements of the dataset, such as cosine similarity \cite{Kavya2015}, Manhattan and Euclidean distances \cite{Mercioni2019, Vagni2021}. In this work, we considered the latter distance because the linkage is performed using the ward method \cite{WARD1963, SZEKELY2005}, which minimizes the sum of the squared distances between cluster centers.

    \subsubsection{Polynomial chaos expansion}\label{subsec:PCE}
    \noindent
    Polynomial chaos expansion (PCE) \cite{Wiener38, Lucor04, LeGratiet17, Sudret2008} is a surrogate\footnote{A surrogate or metamodel is a mathematical model that learns the mapping between the input and the output of a computational model} modeling technique that approximates the behavior/response of complex physics-based models \cite{Berkemeier23, Xiu02, Villegas12}. What makes PCE unique is that it can accelerate tasks such as uncertainty quantification (UQ) and sensitivity analysis \cite{PCA11, NAGEL2020, Luu16}. 
    
    In the model described in \Cref{Sec:CFD}, the conservation equations for momentum, energy, and mass are solved for two input parameters: the substrate temperature ($T_s$) and the chamber pressure ($P$). We express the response of the model as:
	\begin{align}\label{eq:deterM}
		\textbf{y} = \mathcal{M} (\textbf{x}),
	\end{align}
	where $\mathcal{M}(\cdot)$ denotes the CFD model, $\textbf{x} = (\text{x}_{T_s}, \text{x}_{P}) \in \mathbb{R}^{2}$ is the vector containing the values of the two input parameters, and  $\textbf{y} = (\text{y}_1, \text{y}_2, \cdots, \text{y}_N) \in \mathbb{R}^{7 N}$ contains the model responses (see \cref{Fig:Datac}), with $\text{y}_i \in \mathbb{R}^{7}$ for $i = 1, \cdots, N$, and $N$ denoting the dimension of the domain discretization. In the subsequent analysis, we only consider two precursor concentrations as output variables, i.e., $\textbf{y} \in \mathbb{R}^{2 N}$ (see \cref{Sec:Results} for more details).\\
	
    Polynomial chaos computes an approximation of the response, $\bar{\textbf{y}}$, using a set of multi-dimensional polynomials that are orthogonal to the probability density function (PDF) of the input random variables $\textbf{x} = (\text{x}_{T_s}, \text{x}_{P})$. PCE was initially designed for variables following a standard Gaussian distribution, employing Legendre polynomials \cite{Xiu02}. Subsequently, it was extended to other standard statistical distributions (e.g., uniform and beta distributions utilizing Hermite and Jacobi polynomials, respectively \cite{ThB07, LeMaitre12}), or even to arbitrary distributions \cite{Soize04, Wan06}.
	
	Thus, for the vector-value model $\mathcal{M}$, we construct a surrogate model using PCE, denoted as $\mathcal{M}^{PCE}$, in which each component of the model response at each spatial location is represented as follows:
	\begin{align}\label{eq:PCE}
		{\mathcal{M}}^{PCE} (\textbf{\text{x}}) = \sum_{\bm{\alpha} \in \mathbb{N}^{2}} c_{\bm{\alpha}} \bm{\psi}_{\bm{\alpha}}(\textbf{\text{x}}), 
	\end{align}
	where $\bm{\psi}_{\bm{\alpha}}(\cdot)$ represents a multivariate polynomial\footnote{defined as the product of the univariate polynomials: $\bm{\psi}_{\bm{\alpha}}(\textbf{\textit{x}}) = \psi^{(T)}_{\alpha_1}(\textit{x}_{T}) \psi^{(P)}_{\alpha_2}(\textit{x}_{P}).$} that is  orthonormal\footnote{$\mathbb{E}_{\textbf{\textit{x}}}[\bm{\psi}_{\bm{\alpha}}(\textbf{\textit{x}}) \bm{\psi}_{\bm{\beta}}(\textbf{\textit{x}})] = \delta_{\bm{\alpha} \bm{\beta}}$, where $\delta_{\bm{\alpha} \bm{\beta}}$ is the Kronecker delta.} with respect to the PDF of the input parameters,  
	$\bm{\alpha} = (\alpha_1, \alpha_2) \in \mathbb{N}^{2}$ is a multi-index that indicates the polynomial degree of the components of $\bm{\psi}_{\bm{\alpha}}$, and $c_{\bm{\alpha}}$ are the associated coefficients that quantify the relationship between the polynomial expansion ${\mathcal{M}}^{PCE}$ and the computational model ${\mathcal{M}}$. 
	
	In practice, the expansion \eqref{eq:PCE} needs to be truncated to a finite sum ${\mathcal{M}}^{PCE}_d$, where $d \in \mathbb{N}$ is the maximum degree of the polynomials in the expansion:
	\begin{align}\label{eq:PCEtrunq}
		{\mathcal{M}}^{PCE} (\textbf{\text{x}}) \approx {\mathcal{M}}_{d}^{PCE} (\textbf{\text{x}}) + \varepsilon_d = \sum_{\bm{\alpha} \in A_d \subset \mathbb{N}^{2}} {c}_{\bm{\alpha}} \bm{\psi}_{\bm{\alpha}} (\textbf{\text{x}}) + \varepsilon_d, 
	\end{align}
	
	where $\varepsilon_d$ denotes the error introduced due to the truncation of the infinite series, and $A_d$ is a set of multi-indices defined as $A_d = \{\bm{\alpha} \in \mathbb{N}^{2}: |\bm{\alpha}| =  \alpha_1 + \alpha_2 \leq d\}$.

	Next, we estimate the coefficients $c_{\bm{\alpha}}$ while minimizing the error $\varepsilon_d$ \cite{LRS06}. This minimization problem  can be solved, for instance, using nonintrusive methods such as least squares regression (LSR) \cite{Hadigol18}, LASSO regression \cite{Tibshirani94} and least angle regression (LAR) \cite{BLATMAN11}.
	
	Furthermore, to measure the performance of the surrogate model and to avoid overfitting \cite{BLATMAN10}, we use  the \textit{leave-one-out} (LOO) cross-validation error $\varepsilon_{\text{LOO}}$  \cite{saporta11}, which can be implemented as follows:
	\begin{align}\label{eq:LOO}
		\varepsilon_{\text{LOO}} = \nicefrac{\displaystyle \sum_{i = 1}^{m} \left[ \mathcal{M}_d^{PCE} \big(\textbf{\text{x}}^{(i)}\big) - \mathcal{M}_d^{PCE \setminus i} \big(\textbf{\text{x}}^{(i)}\big)\right]^{2}}{\displaystyle\ \text{Var}\left[ \mathcal{M}_d^{PCE} \big(\bm{\mathcal{X}}\big)\right]}, 
	\end{align}
	where $\textbf{\text{x}}^{(i)}$ is an input vector that belongs to the experimental design $\bm{\mathcal{X}}$ (see \cref{Sec:Workflow}), and $\mathcal{M}^{PCE \setminus i}$ is the PCE constructed on the reduced experimental design that excludes $\textbf{\text{x}}^{(i)}$.

    \subsubsection{PCE-based sensitivity analysis (Sobol' indices)}\label{subsubsec:Sobol}
        Sobol' indices \cite{sobol93} are widely used for sensitivity analysis as they provide a robust measure of the input parameters' influence on the output variability. This understanding of parameter importance is particularly valuable in chemical applications \cite{Xie19, santa21}, as it offers crucial physical insights into the process. 
    	
    	Sobol' indices decompose the variance of the model output into the sum of the contributions of individual inputs. As discussed in \cref{subsec:PCE}, the PCE surrogate model \cite{Sudret2008, CRESTAUX09, SUN20} facilitates sensitivity analysis by utilizing the orthogonality of the polynomial basis to estimate the statistical moments of the model response by rewriting expression \eqref{eq:PCEtrunq}, as follows:
	    \begin{align*}
		  {\mathcal{M}}_d^{PCE} (\textbf{\text{x}}) = c_{\textbf{0}} + \sum_{\bm{\alpha} \in A_d \setminus \{\textbf{0}\}} {c}_{\bm{\alpha}} \bm{\psi}_{\bm{\alpha}}(\textbf{\textit{x}}),
	   \end{align*}
        then, its mean and variance are given by \begin{align}\label{eq:m-std}
            \mathbb{E}\left[{\mathcal{M}}_d^{PCE} (\textbf{\text{x}}) \right] = c_{\textbf{0}} \quad \text{ and} \quad  \text{Var}({\mathcal{M}}_d^{PCE} (\textbf{\text{x}})) = \sum_{\bm{\alpha} \in A_d \setminus \{\textbf{0}\} } c_{\bm{\alpha}}^2.
        \end{align}
		
		Thus, we can employ the expressions for the mean and variance above to perform sensitivity analysis using the first-order and total-order Sobol' indices \cite{SUN20}. However, in the following analysis, we are only interested in the total-order Sobol' indices ($S^T$) defined for our two input parameters as
	   \begin{align*}
	       &S^T_{T_s} = \nicefrac{\displaystyle \sum_{\bm{\alpha} \in A_d,\ \alpha_1 \neq 0} c_{\bm{\alpha}}^2\ }{\displaystyle \sum_{\bm{\alpha} \in A_d \setminus \{\textbf{0}\} } c_{\bm{\alpha}}^2},\\
	       &S^T_{P} = \nicefrac{\displaystyle \sum_{\bm{\alpha} \in A_d,\ \alpha_2 \neq 0} c_{\bm{\alpha}}^2\ }{\displaystyle \sum_{\bm{\alpha} \in A_d \setminus \{\textbf{0}\} } c_{\bm{\alpha}}^2}.
	    \end{align*}
		
		These indices are written in terms of PCE coefficients which capture the total influence of the input parameters and their interactions on the variability of the output variables. When $S^T$ is close to one, the contribution of the input parameter is highly significant, whereas values close to zero indicate a negligible influence.

    \subsubsection{Proposed workflow}\label{Sec:scheme}

    In this section, we describe the proposed workflow illustrated in \cref{Fig:workflow}, where each step is enumerated according to the list below that presents a more detailed description:
	
	\begin{enumerate}
		\item[\textit{1.}] \textit{Set an input parameter space and generate the experimental design:} We perform deterministic sampling (DS) \cite{Hessling2013} as a computationally efficient technique for conducting UQ using data from a CFD model \cite{Cutrono2019}. This sampling technique also allows us to incorporate specific input parameter values into the analysis, for which we have experimental data obtained in \cite{14aviziotis2017combined}, facilitating comparison with our results. Thus, given that the input parameters $\textbf{x} = (\text{x}_{T_s}, \text{x}_{P})$ belong to certain production ranges, we uniformly subdivide these ranges with a step size denoted as $h_i$ (see \cref{Tab:inputsG}) and consider all possible combinations of these two input values as experimental design $\bm{\mathcal{X}}$. 
        \begin{table}[ht]
        	\begin{center}
                \caption{Description of the input parameter space.}
        		\begin{tabular}{cccc}
                    \hline
        			Parameter & Description & Range (units) & h$_i$\\
        			\hline
        			$\text{x}_{T_s}$ & Substrate temperature & $[130.0^{\circ}\text{C}, 250.0^{\circ}\text{C}]$ & $0.2^{\circ}\text{C}$\\
        			$\text{x}_{P}$ & Chamber pressure & $[10\ \text{Torr}, 40\ \text{Torr}]$ & $10\ \text{Torr}$\\
        			\hline
        		\end{tabular}
        		\label{Tab:inputsG}
        	\end{center}
        \end{table}
        
        \begin{figure}[pht]
    	   \centering
    	   \includegraphics[width=1\linewidth]{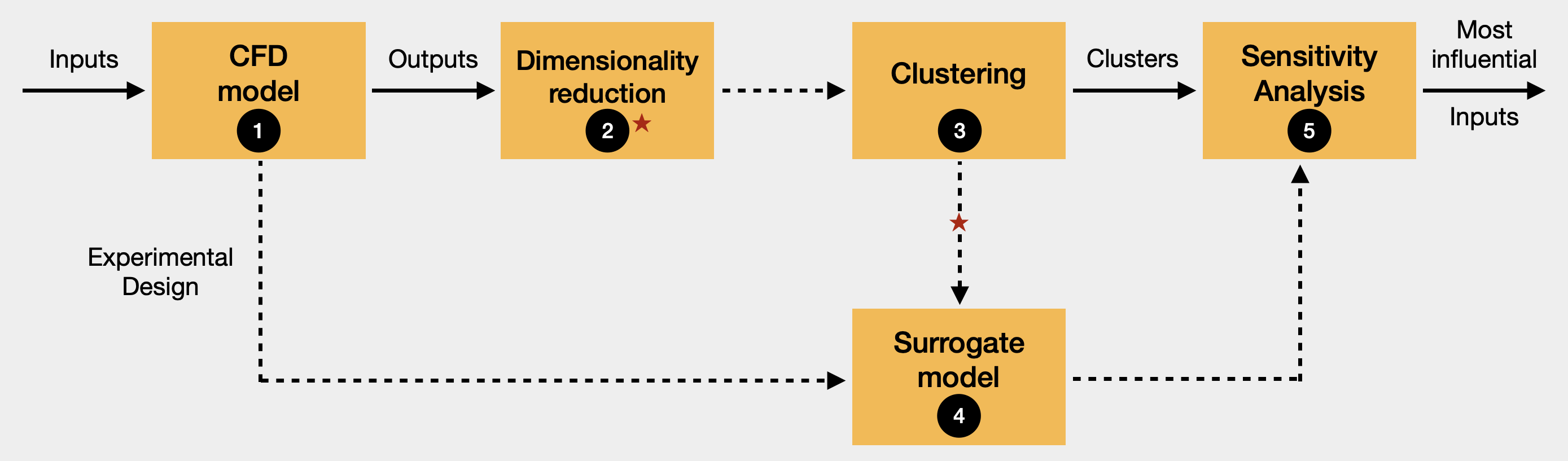}
    	   \caption{Data-driven workflow using machine learning methods to distinguish regimes of the Arrhenius plot  and identify the most influential input parameters. The numbers in each box correspond to the steps detailed in \cref{Sec:scheme}.  The red star on the diagram indicates that it is possible to perform dimensionality reduction at either of the two stages, depending on the size of the dataset.}\label{Fig:workflow}
        \end{figure}

        Using the experimental design, we solve the CFD model and obtain a dataset containing the model responses at all points in the spatial discretization (see \cref{Fig:Datac}).

        \item[\textit{2.}] \textit{Dimensionality reduction:} In this application we consider only the chemical species concentration in an area close to the deposition surface (see \cref{Fig:nodes}), since it is in this area where the critical for deposition phenomena mostly take place; this still constitutes a high-dimensional problem because the model response $\textbf{y} \in \mathbb{R}^{2 N}$, meaning its dimension is twice the number of nodes considered in the spatial discretization of that area. Then, to reduce the dimensionality of data while capturing its variability and to illustrate the clusters in step 3, we apply the PCA method discussed in Section \ref{subsec:PCA}. In the first case, where we reduce the dimensionality of the dataset, we retain the first $n$ principal component modes that describe more than $99\%$ of the data variability, where $n = 3, 5$ and $10$ (See \Cref{Tab:clusters}). In the second case, we retain 10 PCs, but for visualization purposes, we only show the case where we consider 3 PCs.
        
        \item[\textit{3.}] \textit{Clustering and the Arrhenius plot:}  With the low-dimensional representation of the data obtained in the previous step, we apply hierarchical clustering method to identify clusters that correspond to meaningful process regimes.

        \item[\textit{4.}] \textit{Construct the PCE surrogate model:} Based on the distributions followed by the input parameters in each regime, the corresponding families of orthogonal polynomials are considered to construct the PCE model for each of the $n$ retained principal components. The polynomial coefficients are estimated using a Linear Regression method, as discussed in Section \ref{subsec:PCE}. In addition, we analyze the evolution of the LOO-error, introduced in \eqref{eq:LOO}, for each polynomial degree, until reaching an optimal degree for which this error is sufficiently small (lower than order $O(-3)$).

        \item[\textit{5.}] \textit{PCE-based uncertainly quantification and sensitivity analysis:} At this stage, we use the PCE surrogate model to directly, and accurately, estimate the moments of the model responses based on the polynomial coefficients. These moments are used for sensitivity analysis through the Sobol' indices, which explain the influence of input variability on the model responses. This allows us to identify important input parameters, as explained in \Cref{subsubsec:Sobol}.
    \end{enumerate} 
    
    Steps \textit{2}, \textit{3}, \textit{4}, and \textit{5} were implemented in Python using the Sklearn and UQpy packages \cite{SKL, UQpy}.

\section{Results and discussion}\label{Sec:Results}
	Inspired by the Arrhenius plot shown in Fig. \ref{Fig:Arrh}, the results and analysis presented below are based on the concentration values of two different species: Fe(CO)$_5$ and Fe(CO), as calculated by the CFD model described in Section \ref{Sec:CFD} for the parameter values reported in Table \ref{Tab:inputsG}. These species were chosen because their decomposition has the lowest activation energy (see \Cref{Tab:surface-reactions}). This implies that their contribution to the overall deposition rate, determined by an Arrhenius model, is considerably more significant than those obtained with other species.
	
\begin{figure}[!ht]
 \centering
    \includegraphics[width=0.2\linewidth]{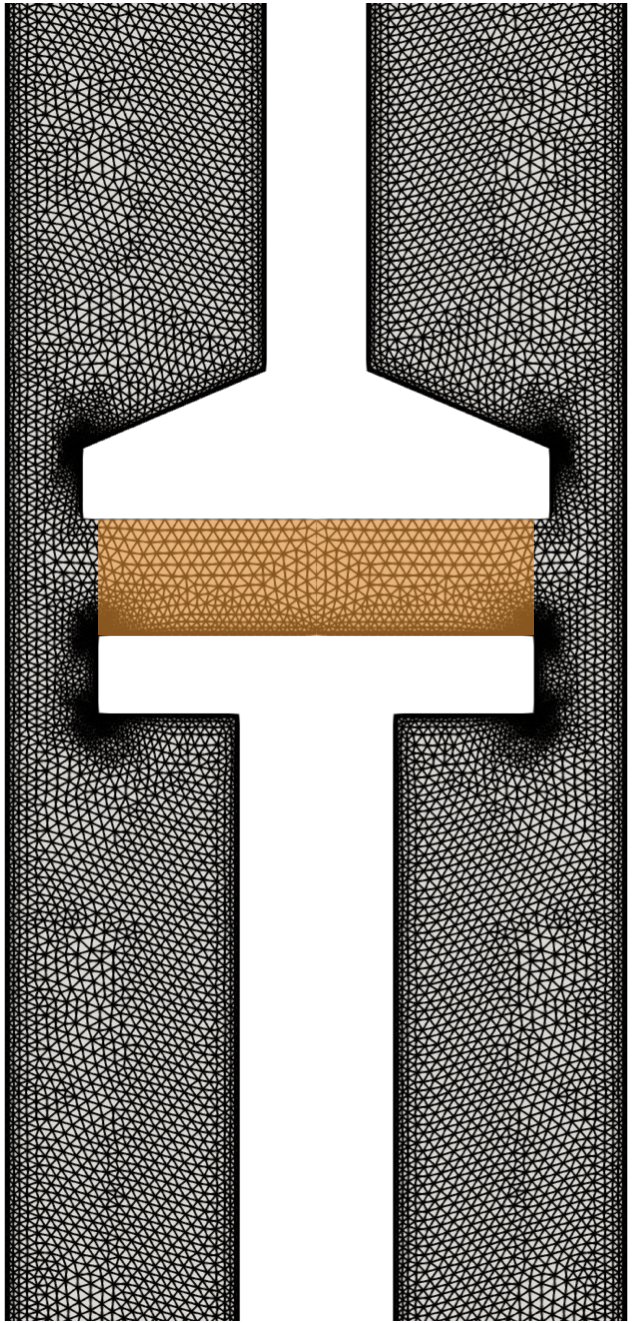}
		\caption{Nodes of the spatial discretization (all points) considered to solve the computational model. Data from the nodes close to the deposition surface (orange area) are used in the data-driven analysis.}\label{Fig:nodes}
	\end{figure}

	Moreover, in this application it makes sense to consider data in the proximity to the deposition surface, shown in orange colour in \cref{Fig:nodes}), since the species concentration in that area determines the deposition rate and its distribution.
	
    \subsection{Dimensionality reduction}

    For this initial stage, the area of interest is discretized by a computational mesh with $828$ nodes, where the species concentration are calculated. The size of the data scales with the discretization of the physical domain, making dimensionality reduction necessary to keep the meaningful information in a low-dimensional representation, which will be used to build a more efficient surrogate model.

    \begin{figure}[!pht]
		\centering
		\includegraphics[width = 0.7\linewidth]{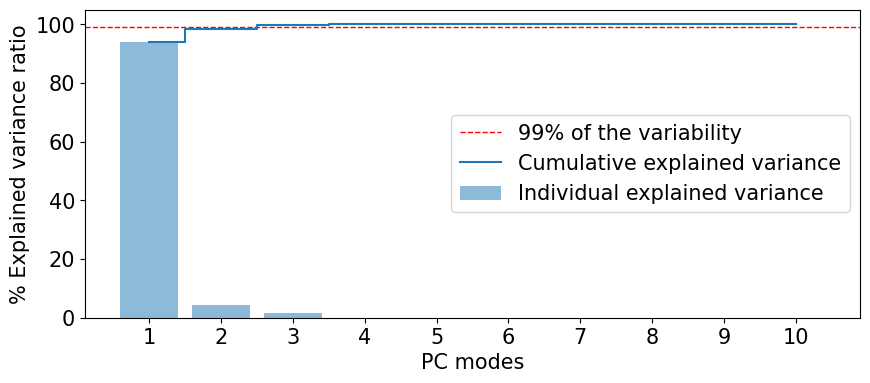}
		\caption{Evolution of the individual (blue bars) and cumulative (blue line) explained variance as a function of the number of retained PC modes. In addition, a minimum threshold (red dashed line) is set at $99\%$ of the total variability.}\label{figPCA51020}
	\end{figure}
    
    Precisely, PCA allows us to quantify the variability explained by each direction in which the data is projected. We repeat this exercise for different numbers of PCs as we want to examine the potential effects of dimensionality reduction in our subsequent analysis. Here we have considered retaining 3, 5, and 10 PCs. The evolution of the explained variability as a function of the number of retained PCs is shown in \cref{figPCA51020}, where we observe that it is sufficient to retain three principal components for the established threshold to be exceeded with $99\%$. 


    \begin{figure}[!pht]
        \centering
        \includegraphics[width=0.7\linewidth]{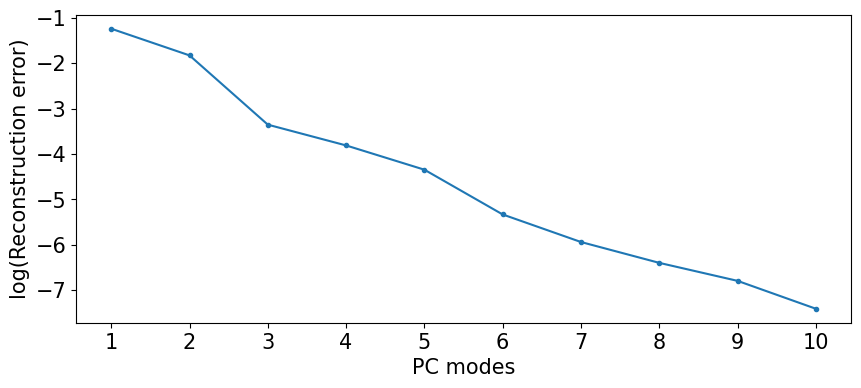}
        \caption{Evolution of the reconstruction error (blue line) in a logarithmic scale as a function of the number of retained PC modes.}
        \label{fig:figRE51020}
        \end{figure}

    On the other hand, when observing the behavior of the reconstruction error presented in \cref{fig:figRE51020} as a semi-log plot, it becomes clear that the error drops to $O(-3)$ with 3 PCs, and it drops below $O(-6)$ when at least 8 PCs are considered. This suggests that in this application at most 10 PCs are required to accurately reconstruct the data.

\subsection{Clustering: identification of deposition regimes}\label{subc:clus}

    In the analysis that follows, each chamber pressure $P$ is treated separately, by implementing Hierarchical Clustering (refer to \cref{subsec:Cluster}) to data collected in each value of chamber pressure. In each case, three entirely disjoint clusters are identified. 
    
\begin{figure}[ht]
		\centering
		\includegraphics[width=0.7 \linewidth]{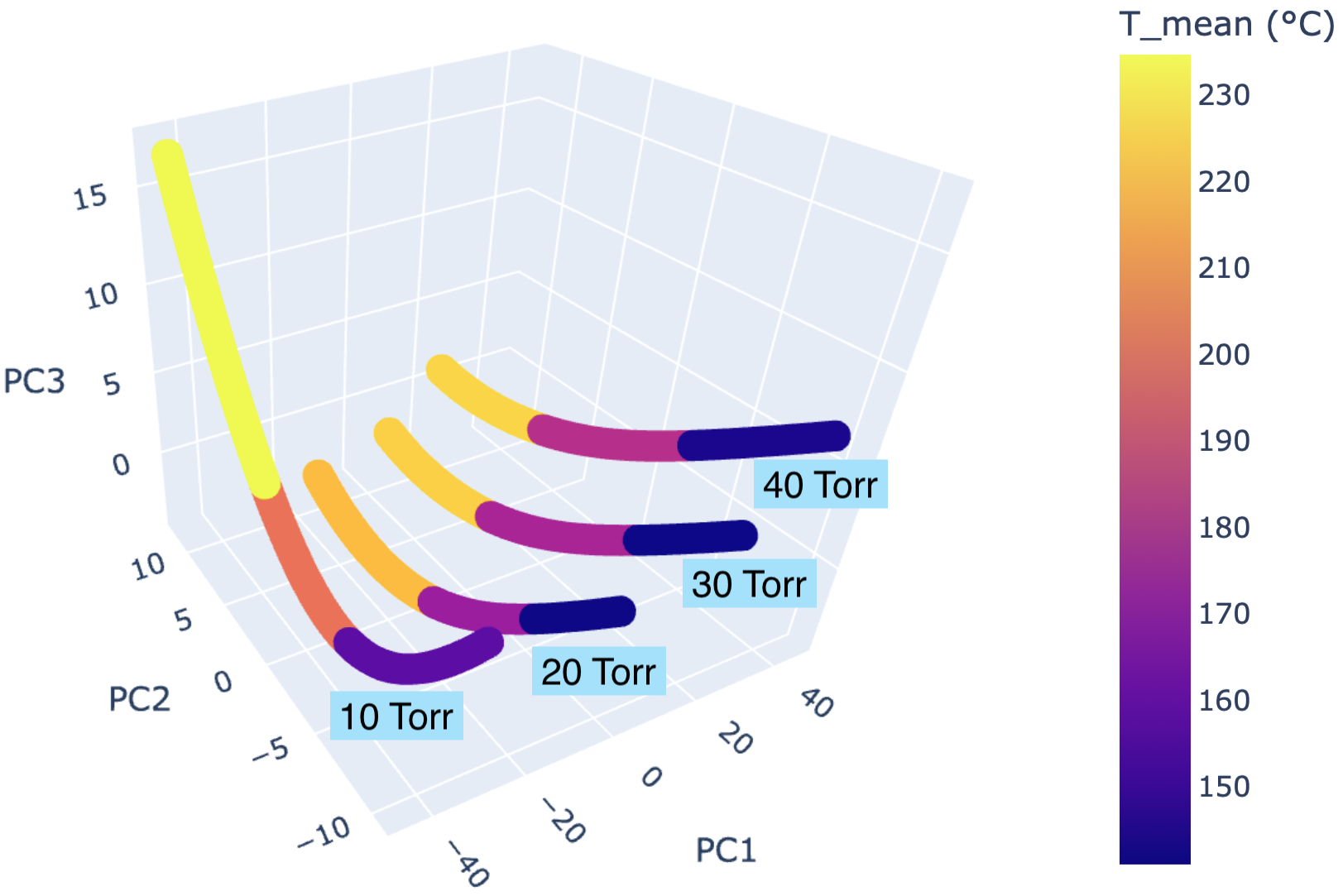}
		\caption{The clusters are projected in a three-dimensional space using PCA, for each value of $P =10, 20, 30$ and $40$ Torr. They are colored according to the average substrate temperature $T_s$ in °C, following the temperature scale shown on the right, with blue representing the lowest temperature and yellow representing the highest ones.}\label{fig:clusters3D}
	\end{figure}

	
	To visualize the clusters identified the data is projected on a lower-dimensional space, identified by the PCA method described in \cref{subsec:PCA}, as illustrated in \cref{fig:clusters3D}. 
 Within each group corresponding to different values of chamber pressure $P$, we can identify three separate clusters, each represented by a distinct color, determined by the average temperature of the substrate from which the values of each cluster are derived. It is worth noting that the clustering process focused solely on species concentrations, making it entirely oblivious of the input parameters of the computational model, i.e. the operational parameters $T_s$ and $P$. This practical approach highlights the capability of HC to efficiently identify clusters of interest using only computational experiments. 
 
    As discussed in the previous section, we explore how the clustering results change as we increase the number of principal components, as depicted in \Cref{Tab:clusters}. We examine cases with 3, 5, and 10 PCs, revealing the convergence of temperature ranges associated with regimes for each pressure value, which means that the clusters we obtained are insensitive to the number of principal components considered. Specifically, the results presented in \Cref{Tab:clusters}, corresponding to $P=30$ Torr and $P=40$ Torr, indicate that the range associated with the intermediate regime (in the green rows) shows no discernible changes beyond the results obtained when considering $3$ PCs or more. For $P=20$ Torr, this range narrows for $5$ PCs, but no further changes occur beyond that point. For the last case  ($P=10$ Torr), we observe that this intermediate range remains unchanged for a small number of PCs, but when considering 10 PCs, it  tightens up slightly. Therefore, the subsequent analyses focuses on the scenario where $10$ PCs are taken into account, ensuring the elimination of potential effects resulting from dimensionality reduction. Moreover, intervals corresponding to $10$ PCs are colored in the fourth column of \Cref{Tab:clusters} to identify them with each regime.
    \begin{table}[ht]
        \centering
        \caption{Temperature ranges associated with the clusters identified in \cref{fig:clusters3D} for each value of $P$ arranged in ascending order.}
        {\scriptsize \begin{tabular}{cccccccccccc}
            \cline{3-4} \cline{6-7} \cline{9-10} \cline{12-12}
            & \multicolumn{1}{c|}{} & \multicolumn{1}{c|}{\cellcolor[HTML]{D2D2D2}T$_{\text{min}}$}  & \multicolumn{1}{c|}{\cellcolor[HTML]{D2D2D2}T$_{\text{max}}$}  & \multicolumn{1}{c|}{} & \multicolumn{1}{c|}{\cellcolor[HTML]{D2D2D2}T$_{\text{min}}$}  & \multicolumn{1}{c|}{\cellcolor[HTML]{D2D2D2}T$_{\text{max}}$}  & \multicolumn{1}{c|}{} & \multicolumn{1}{c|}{\cellcolor[HTML]{D2D2D2}T$_{\text{min}}$}  & \multicolumn{1}{c|}{\cellcolor[HTML]{D2D2D2}T$_{\text{max}}$}  & \multicolumn{1}{c|}{} & \multicolumn{1}{c|}{\cellcolor[HTML]{D2D2D2}P}                    \\ \cline{1-1} \cline{3-4} \cline{6-7} \cline{9-10} \cline{12-12} 
            \multicolumn{1}{|c|}{\cellcolor[HTML]{FFC702}Cluster 3}                        & \multicolumn{1}{c|}{} & \multicolumn{1}{c|}{203.4 °C}                         & \multicolumn{1}{c|}{250.0 °C}                         & \multicolumn{1}{c|}{} & \multicolumn{1}{c|}{203.4 °C}                         & \multicolumn{1}{c|}{250.0 °C}                         & \multicolumn{1}{c|}{} & \multicolumn{1}{c|}{\cellcolor[HTML]{FFCE93}203.4 °C} & \multicolumn{1}{c|}{\cellcolor[HTML]{FFCE93}250.0 °C} & \multicolumn{1}{c|}{} & \multicolumn{1}{c|}{\cellcolor[HTML]{96FFFB}}                     \\ \cline{1-1} \cline{3-4} \cline{6-7} \cline{9-10}
            \multicolumn{1}{|c|}{\cellcolor[HTML]{BD19D9}{\color[HTML]{FFFFFF} Cluster 2}} & \multicolumn{1}{c|}{} & \multicolumn{1}{c|}{\cellcolor[HTML]{B8F69B}157.4 °C} & \multicolumn{1}{c|}{\cellcolor[HTML]{B8F69B}203.2 °C} & \multicolumn{1}{c|}{} & \multicolumn{1}{c|}{\cellcolor[HTML]{B8F69B}157.4 °C} & \multicolumn{1}{c|}{\cellcolor[HTML]{B8F69B}203.2 °C} & \multicolumn{1}{c|}{} & \multicolumn{1}{c|}{\cellcolor[HTML]{B8F69B}157.4 °C} & \multicolumn{1}{c|}{\cellcolor[HTML]{B8F69B}203.2 °C} & \multicolumn{1}{c|}{} & \multicolumn{1}{c|}{\cellcolor[HTML]{96FFFB}}                     \\ \cline{1-1} \cline{3-4} \cline{6-7} \cline{9-10}
            \multicolumn{1}{|c|}{\cellcolor[HTML]{00009B}{\color[HTML]{FFFFFF} Cluster 1}} & \multicolumn{1}{c|}{} & \multicolumn{1}{c|}{130.0 °C}                         & \multicolumn{1}{c|}{157.2 °C}                         & \multicolumn{1}{c|}{} & \multicolumn{1}{c|}{130.0 °C}                         & \multicolumn{1}{c|}{157.2 °C}                         & \multicolumn{1}{c|}{} & \multicolumn{1}{c|}{\cellcolor[HTML]{A5EFFF}130.0 °C} & \multicolumn{1}{c|}{\cellcolor[HTML]{A5EFFF}157.2 °C} & \multicolumn{1}{c|}{} & \multicolumn{1}{c|}{\multirow{-3}{*}{\cellcolor[HTML]{96FFFB}40}} \\ \cline{1-1} \cline{3-4} \cline{6-7} \cline{9-10} \cline{12-12} 
            &  &  &  &  &   &  &  &  &  & \\\cline{1-1} \cline{3-4} \cline{6-7} \cline{9-10} \cline{12-12} 
            \multicolumn{1}{|c|}{\cellcolor[HTML]{FFC702}Cluster 3}                        & \multicolumn{1}{c|}{} & \multicolumn{1}{c|}{192.2 °C}                         & \multicolumn{1}{c|}{250.0 °C}                         & \multicolumn{1}{c|}{} & \multicolumn{1}{c|}{192.2 °C}                         & \multicolumn{1}{c|}{250.0 °C}                         & \multicolumn{1}{c|}{} & \multicolumn{1}{c|}{\cellcolor[HTML]{FFCE93}192.2 °C} & \multicolumn{1}{c|}{\cellcolor[HTML]{FFCE93}250.0 °C} & \multicolumn{1}{c|}{} & \multicolumn{1}{c|}{\cellcolor[HTML]{96FFFB}}                     \\ \cline{1-1} \cline{3-4} \cline{6-7} \cline{9-10}
            \multicolumn{1}{|c|}{\cellcolor[HTML]{BD19D9}{\color[HTML]{FFFFFF} Cluster 2}} & \multicolumn{1}{c|}{} & \multicolumn{1}{c|}{\cellcolor[HTML]{B8F69B}153.8 °C} & \multicolumn{1}{c|}{\cellcolor[HTML]{B8F69B}192.0 °C} & \multicolumn{1}{c|}{} & \multicolumn{1}{c|}{\cellcolor[HTML]{B8F69B}153.8 °C} & \multicolumn{1}{c|}{\cellcolor[HTML]{B8F69B}192.0 °C} & \multicolumn{1}{c|}{} & \multicolumn{1}{c|}{\cellcolor[HTML]{B8F69B}153.8 °C} & \multicolumn{1}{c|}{\cellcolor[HTML]{B8F69B}192.0 °C} & \multicolumn{1}{c|}{} & \multicolumn{1}{c|}{\cellcolor[HTML]{96FFFB}}                     \\ \cline{1-1} \cline{3-4} \cline{6-7} \cline{9-10}
            \multicolumn{1}{|c|}{\cellcolor[HTML]{00009B}{\color[HTML]{FFFFFF} Cluster 1}} & \multicolumn{1}{c|}{} & \multicolumn{1}{c|}{130.0 °C}                         & \multicolumn{1}{c|}{153.6 °C}                         & \multicolumn{1}{c|}{} & \multicolumn{1}{c|}{130.0 °C}                         & \multicolumn{1}{c|}{153.6 °C}                         & \multicolumn{1}{c|}{} & \multicolumn{1}{c|}{\cellcolor[HTML]{A5EFFF}130.0 °C} & \multicolumn{1}{c|}{\cellcolor[HTML]{A5EFFF}153.6 °C} & \multicolumn{1}{c|}{} & \multicolumn{1}{c|}{\multirow{-3}{*}{\cellcolor[HTML]{96FFFB}30}} \\ \cline{1-1} \cline{3-4} \cline{6-7} \cline{9-10} \cline{12-12} 
            &  &  &  &  &   &  &  &  &  & \\ \cline{1-1} \cline{3-4} \cline{6-7} \cline{9-10} \cline{12-12} 
            \multicolumn{1}{|c|}{\cellcolor[HTML]{FFC702}Cluster 3}                        & \multicolumn{1}{c|}{} & \multicolumn{1}{c|}{221.2 °C}                         & \multicolumn{1}{c|}{250.0 °C}                         & \multicolumn{1}{c|}{} & \multicolumn{1}{c|}{191.0 °C}                         & \multicolumn{1}{c|}{250.0 °C}                         & \multicolumn{1}{c|}{} & \multicolumn{1}{c|}{\cellcolor[HTML]{FFCE93}191.0 °C} & \multicolumn{1}{c|}{\cellcolor[HTML]{FFCE93}250.0 °C} & \multicolumn{1}{c|}{} & \multicolumn{1}{c|}{\cellcolor[HTML]{96FFFB}}                     \\ \cline{1-1} \cline{3-4} \cline{6-7} \cline{9-10}
            \multicolumn{1}{|c|}{\cellcolor[HTML]{BD19D9}{\color[HTML]{FFFFFF} Cluster 2}} & \multicolumn{1}{c|}{} & \multicolumn{1}{c|}{\cellcolor[HTML]{B8F69B}170.2 °C} & \multicolumn{1}{c|}{\cellcolor[HTML]{B8F69B}221.0 °C} & \multicolumn{1}{c|}{} & \multicolumn{1}{c|}{\cellcolor[HTML]{B8F69B}152.6 °C} & \multicolumn{1}{c|}{\cellcolor[HTML]{B8F69B}190.8 °C} & \multicolumn{1}{c|}{} & \multicolumn{1}{c|}{\cellcolor[HTML]{B8F69B}152.6 °C} & \multicolumn{1}{c|}{\cellcolor[HTML]{B8F69B}190.8 °C} & \multicolumn{1}{c|}{} & \multicolumn{1}{c|}{\cellcolor[HTML]{96FFFB}}                     \\ \cline{1-1} \cline{3-4} \cline{6-7} \cline{9-10}
            \multicolumn{1}{|c|}{\cellcolor[HTML]{00009B}{\color[HTML]{FFFFFF} Cluster 1}} & \multicolumn{1}{c|}{} & \multicolumn{1}{c|}{130.0 °C}                         & \multicolumn{1}{c|}{170.0 °C}                         & \multicolumn{1}{c|}{} & \multicolumn{1}{c|}{130.0 °C}                         & \multicolumn{1}{c|}{152.4 °C}                         & \multicolumn{1}{c|}{} & \multicolumn{1}{c|}{\cellcolor[HTML]{A5EFFF}130.0 °C} & \multicolumn{1}{c|}{\cellcolor[HTML]{A5EFFF}152.4 °C} & \multicolumn{1}{c|}{} & \multicolumn{1}{c|}{\multirow{-3}{*}{\cellcolor[HTML]{96FFFB}20}} \\ \cline{1-1} \cline{3-4} \cline{6-7} \cline{9-10} \cline{12-12} 
            &  &  &  &  &   &  &  &  &  & \\ \cline{1-1} \cline{3-4} \cline{6-7} \cline{9-10} \cline{12-12} 
            \multicolumn{1}{|c|}{\cellcolor[HTML]{FFEC00}Cluster 3}                        & \multicolumn{1}{c|}{} & \multicolumn{1}{c|}{219.2 °C}                         & \multicolumn{1}{c|}{250.0 °C}                         & \multicolumn{1}{c|}{} & \multicolumn{1}{c|}{219.2°C}                          & \multicolumn{1}{c|}{250.0 °C}                         & \multicolumn{1}{c|}{} & \multicolumn{1}{c|}{\cellcolor[HTML]{FFCE93}219.2 °C} & \multicolumn{1}{c|}{\cellcolor[HTML]{FFCE93}250.0 °C} & \multicolumn{1}{c|}{} & \multicolumn{1}{c|}{\cellcolor[HTML]{96FFFB}}                     \\ \cline{1-1} \cline{3-4} \cline{6-7} \cline{9-10}
            \multicolumn{1}{|c|}{\cellcolor[HTML]{FBAB0E}Cluster 2}                        & \multicolumn{1}{c|}{} & \multicolumn{1}{c|}{\cellcolor[HTML]{B8F69B}184.2 °C} & \multicolumn{1}{c|}{\cellcolor[HTML]{B8F69B}219.0 °C} & \multicolumn{1}{c|}{} & \multicolumn{1}{c|}{\cellcolor[HTML]{B8F69B}184.2 °C} & \multicolumn{1}{c|}{\cellcolor[HTML]{B8F69B}219.0 °C} & \multicolumn{1}{c|}{} & \multicolumn{1}{c|}{\cellcolor[HTML]{B8F69B}183.4° C} & \multicolumn{1}{c|}{\cellcolor[HTML]{B8F69B}219.0 °C} & \multicolumn{1}{c|}{} & \multicolumn{1}{c|}{\cellcolor[HTML]{96FFFB}}                     \\ \cline{1-1} \cline{3-4} \cline{6-7} \cline{9-10}
            \multicolumn{1}{|c|}{\cellcolor[HTML]{CB00EE}{\color[HTML]{FFFFFF} Cluster 1}} & \multicolumn{1}{c|}{} & \multicolumn{1}{c|}{130.0 °C}                         & \multicolumn{1}{c|}{184.0 °C}                         & \multicolumn{1}{c|}{} & \multicolumn{1}{c|}{130.0 °C}                         & \multicolumn{1}{c|}{184.0°C}                          & \multicolumn{1}{c|}{} & \multicolumn{1}{c|}{\cellcolor[HTML]{A5EFFF}130.0 °C} & \multicolumn{1}{c|}{\cellcolor[HTML]{A5EFFF}183.2 °C} & \multicolumn{1}{c|}{} & \multicolumn{1}{c|}{\multirow{-3}{*}{\cellcolor[HTML]{96FFFB}10}} \\ \cline{1-1} \cline{3-4} \cline{6-7} \cline{9-10} \cline{12-12} 
            &  &  &  &  &   &  &  &  &  & \\ \cline{3-4} \cline{6-7} \cline{9-10}
            & \multicolumn{1}{c|}{} & \multicolumn{2}{c|}{\cellcolor[HTML]{67FD9A}3 PCs}                                                            & \multicolumn{1}{c|}{} & \multicolumn{2}{c|}{\cellcolor[HTML]{67FD9A}5 PCs}                                                            & \multicolumn{1}{c|}{} & \multicolumn{2}{c|}{\cellcolor[HTML]{67FD9A}10 PCs}                                                           &                       &                   \\ \cline{3-4} \cline{6-7} \cline{9-10}
        \end{tabular}}
		\label{Tab:clusters}
	\end{table}
    
    We then proceed to compare the temperature range of the clusters for $P=10$ Torr (refer to the cluster block at the bottom of \Cref{Tab:clusters}) to the regimes of the experimentally determined Arrhenius plot (cf. \cref{Fig:Arrh}).
 Indeed, upon analyzing these findings presented at the bottom of \Cref{Tab:clusters}, it becomes apparent that Cluster 1 coincides with the temperature range for the reaction-limited regime, i.e. $130.0-183.2$ °C. Additionally, Cluster 2 coincides with the transition regime, which occurs for $183.4-219.0$ °C, while Cluster 3 corresponds to the diffusion-limited regime, encompassing the span of $219.2-250.0$ °C. 
 
\begin{figure}[ht!]
 	  \centering
	    \includegraphics[width=0.7 \linewidth]{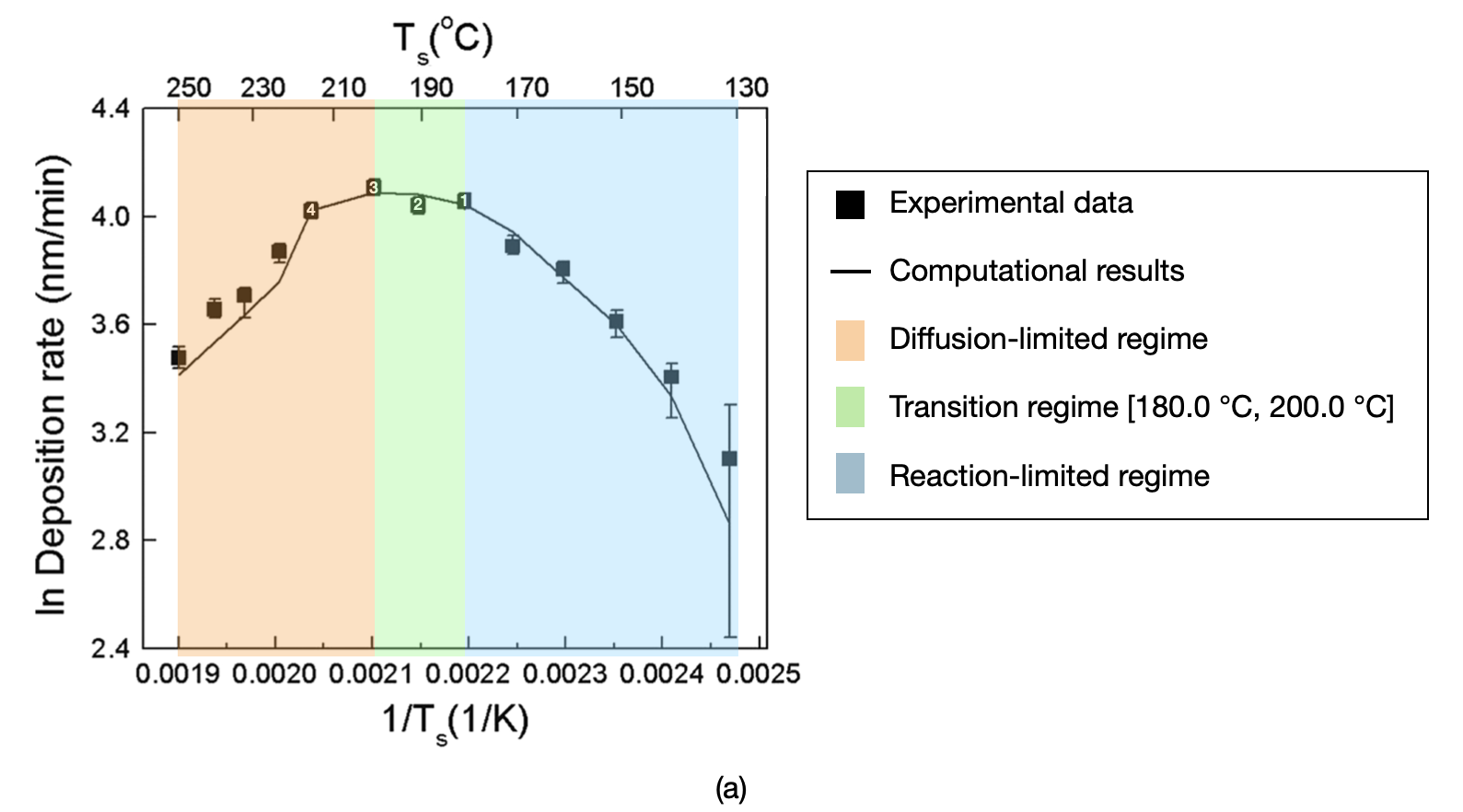}
        \includegraphics[width=0.7 \linewidth]{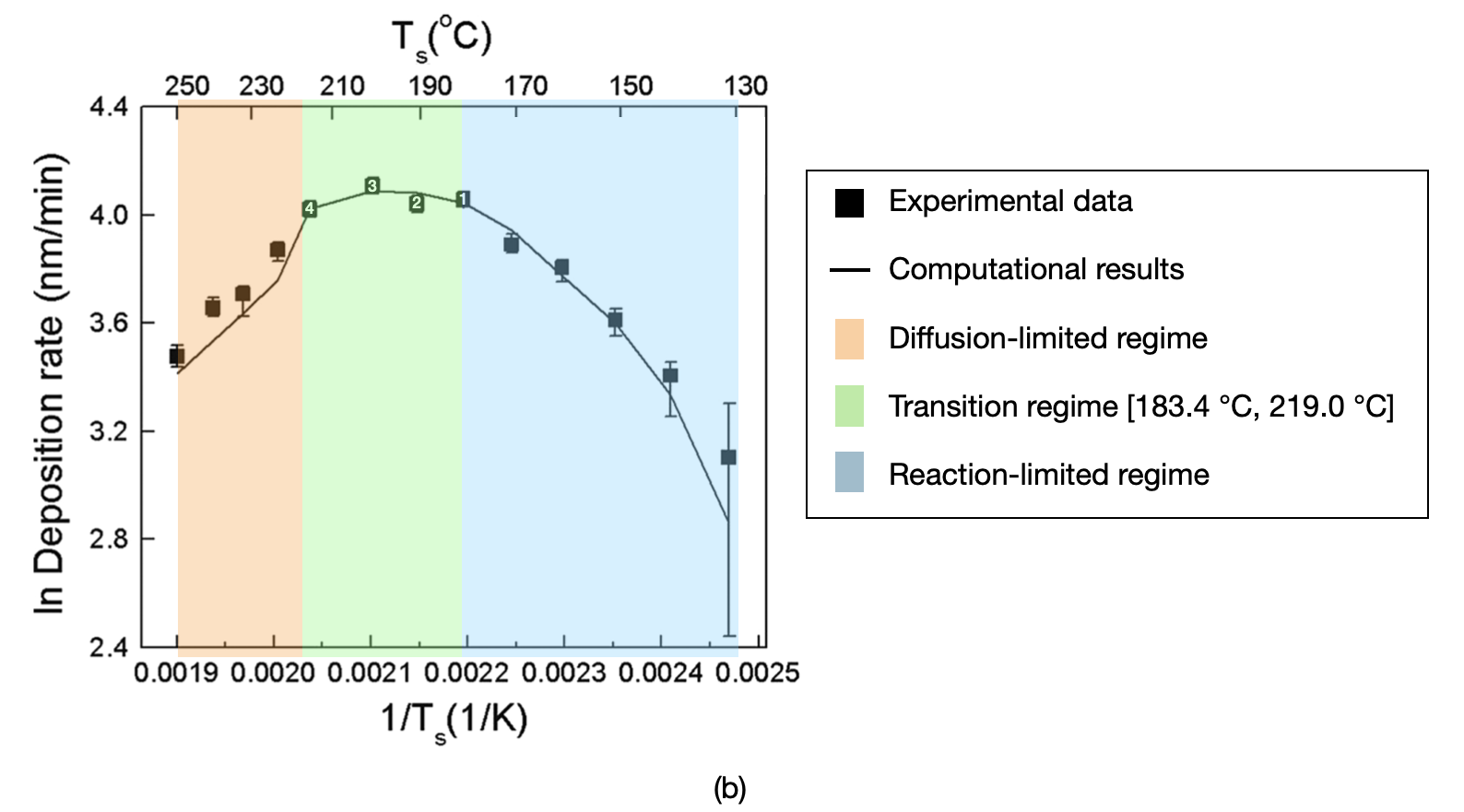}
	   \caption{The Arrhenius plot of the Fe-CVD reactor for $P = 10$ Torr. Three regions have been colored: reaction-limited regime (low temperatures) in blue, transition regime in green, and diffusion-limited regime (high temperatures) in orange. These regions were colored according to (a) the results presented in \cite{14aviziotis2017combined} and (b) temperature ranges associated with the clusters obtained in this study.}\label{Fig:P10}
    \end{figure}

 Notably, these values indicate a transition regime extending to slightly higher temperatures compared to the results presented in \cite{14aviziotis2017combined}. Specifically, if we rely visually on the Arrhenius plots shown in \cref{Fig:P10}, we observe four numbered points aligned at the top of the curve, which could be interpreted as region where the reaction and the species diffusion rates become similar. Additionally, three disjoint and colored regions corresponding to the three previously defined regimes appear in each case.  \cref{Fig:P10}(a) displays the Arrhenius plot ($P=10$ Torr) where these regions are shown according to findings in \cite{14aviziotis2017combined}, with the intermediate regime (in green) encompassing points 1, 2, and 3, extending up to $200$ °C. This is because, in \cite{14aviziotis2017combined}, some changes in the crystalline structure were observed at that temperature, urging the authors to classify the fourth point (at $T=220$ °C) in the diffusion-limited regime rather than in the intermediate regime. In contrast, the results obtained with hierarchical clustering, shown  \cref{Fig:P10}(b), indicate that the intermediate regime (in green) extends a bit further to include point 4. As mentioned, point 4 aligns with the other three points, suggesting its independence from temperature. Therefore, it seems reasonable to include it in this region, as our results indicate.
 
    For higher pressures, the clustering results indicate a marked shift of the behavior toward lower temperatures. Namely, for the case of $P = 20$ Torr, we observe that the temperature ranges associated with the clusters are as follows: Cluster 1, linked to the reaction-limited regime, spans from $130.0-152.4$ °C. Cluster 2, associated with the transition regime, falls between $152.6-190.8$ °C, while Cluster 3, representing the diffusion-limited regime, ranges from $191.0-250.0$ °C. Similarly, at $P = 30$ Torr, we obtain the following results: the reaction-limited regime covers temperatures between $130.0-153.6$ °C, and the transition regime covers temperatures between $153.8-192.0$ °C. The diffusion-limited regime remains between $192.2-250.0$ °C. For $P = 40$ Torr, we observe temperature ranges analogous to the previous cases: the reaction-limited regime is between $130.0-157.2$ °C, the transition regime spans in the range $157.4-203.2$ °C, and the diffusion-limited regime ranges between $203.4-250.0$ °C.

    For higher chamber pressures $P \geq 20$ Torr, we have experimental data from only a single temperature, $T_s = 180$ °C. According to \cite{14aviziotis2017combined}, at this temperature and $P=40$ Torr, the deposition rate value is $1/8$ of the deposition rate at the same temperature but at $P=10$ Torr. This implies that the deposition rate significantly decreases at higher pressures, as will be discussed later. Our computational model, also follows this trend, finding that the deposition rate at $P=40$ Torr is 1/7 of the value at $P=10$ Torr. This is further confirmation of the accuracy of the CFD model and the clustering algorithm, at parameter values that are outside the training set.


    \subsection{PCE surrogate model}
    
    Proceeding to surrogate modeling, we employ Polynomial Chaos Expansion (PCE) as discussed in \cref{subsec:PCE}. We formulate surrogates model for each position in the spatial distribution in the area of interest and for each regime of the Arrhenius plot.
    \begin{table}[ht!]
		\begin{center}
            \caption{Distributions of the parameter $T_s$ for building PCE models in each regime.}\label{Tab:distributions1020}
			\begin{tabular}{ccc}
				\hline Regime & Distribution \\
				\hline
				Diffusion-limited & $\mathcal{U}([T_{s\text{\_min\_diffusion}}, T_{s\text{\_max\_diffusion}}])$ \\
				Transition & $\mathcal{U}([T_{s\text{\_min\_transition}}, T_{s\text{\_max\_transition}}])$ \\
				Reaction-limited & $\mathcal{U}([T_{s\text{\_min\_reaction}}, T_{s\text{\_max\_reaction}}])$ \\
				\hline
			\end{tabular}
		\end{center}
	\end{table}
    
    Furthermore, since the process relies on $\text{x}_{T_s}$ and $\text{x}_P$, and both parameters are treated as random variables for constructing the PCE model, it is essential to consider their probabilistic information. However, as the distributions of these random variables are unknown, we assume they follow uniform distributions. More precisely, $\text{x}_P$ follows the distribution $\mathcal{U}([10.0, 40.0])$ across all regimes. Meanwhile, $\text{x}_{T_s}$ follows a uniform distribution defined in a specific interval for each regime and each pressure (according to the results presented in \cref{Tab:clusters}), as detailed in \cref{Tab:distributions1020}.
 
	Under these assumptions, we construct surrogate models usisng Legendre polynomials associated with the uniform distribution. Their coefficients are computed employing the Least Squares Regression (LSR) method (see \cref{subsec:PCE}).
    \begin{figure}[!ht] 
        \centering
        \includegraphics[width=0.6\linewidth]{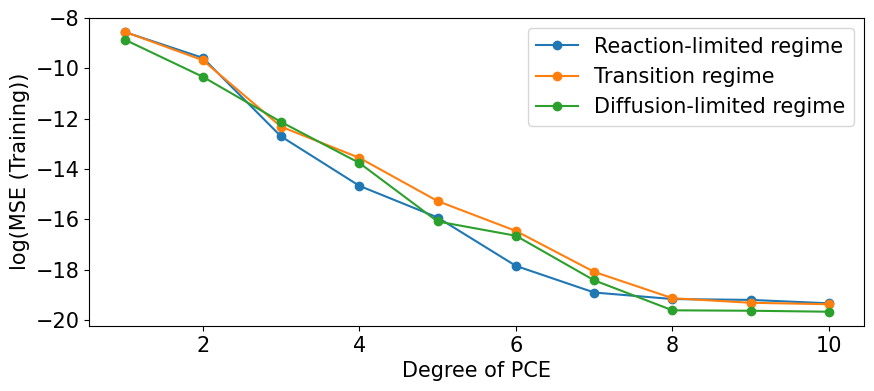} 
        \includegraphics[width=0.6\linewidth]{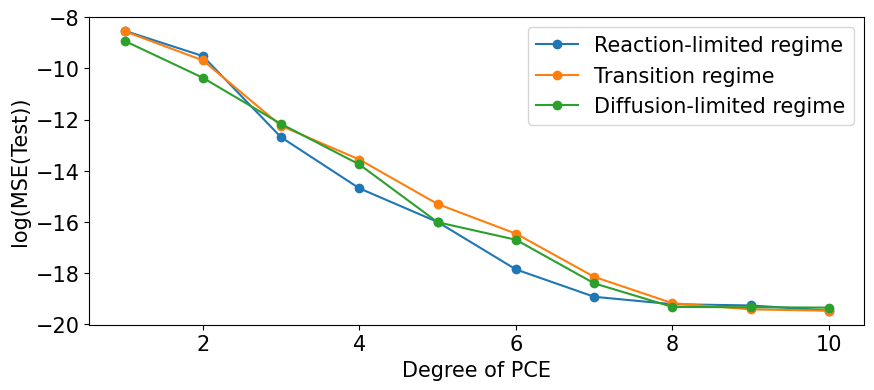} 
        \caption{Evolution of the training (at the top) and test (at the bottom) as a function of the degree of PCE for the reaction-limited regime, transition regime, and diffusion-limited regime.}\label{train-1020}
	\end{figure}
    We assess the accuracy of this approach by studying the training and test errors shown in \cref{train-1020}. These results indicate accurate predictions of spatial distribution concentrations, in every regime of the Arrhenius plot. In each case, we achieve a low-order for both errors, confirming the high accuracy of the PCE models. Based on their evolution, the maximum degree for the surrogate approximation is set at 3.
    
	Having established that PCE can produce accurate surrogate models, we can exploit this opportunity further and extract crucial information to quantify the inputs effects propagated through the surrogate model in terms of Sobol' indices, as discussed in the following section.
 
    \subsection{PCE-based uncertainty quantification}
\begin{figure}[!ht] 
        \centering
            \includegraphics[width=0.50\linewidth]{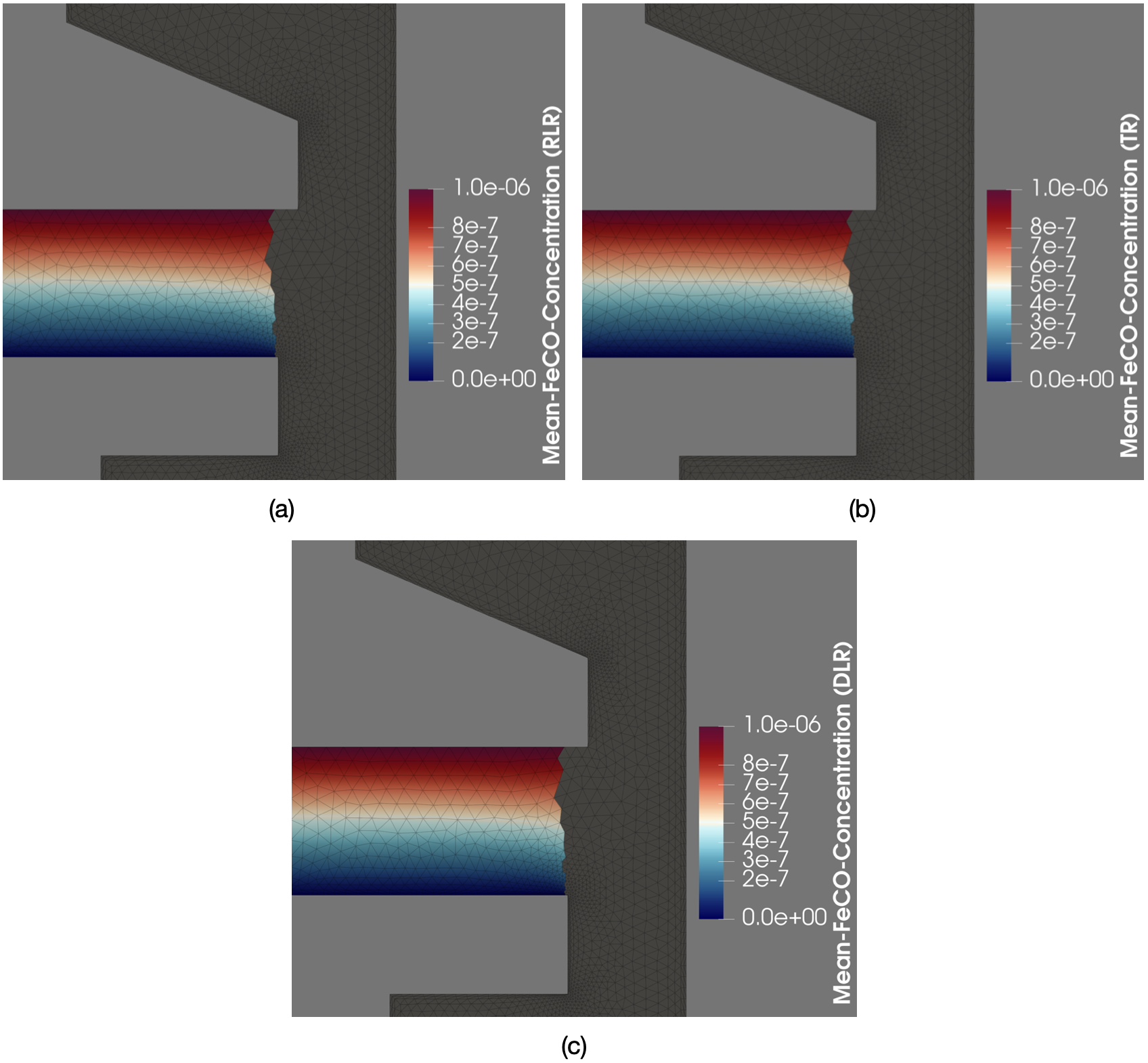}
		    \caption{Spatial distribution over the region near the deposition surface of the mean for Fe(CO) concentrations (P = 10 Torr) in (a) the reaction-limited regime, (b) the transition regime, and (c) the diffusion-limited regime. The color scale on the right indicates darker colors for smaller quantities and lighter colors for larger quantities.
        }
        \label{M-FeCO}
	\end{figure}

\begin{figure}[!ht] 
        \centering
            \includegraphics[width=0.50\linewidth]{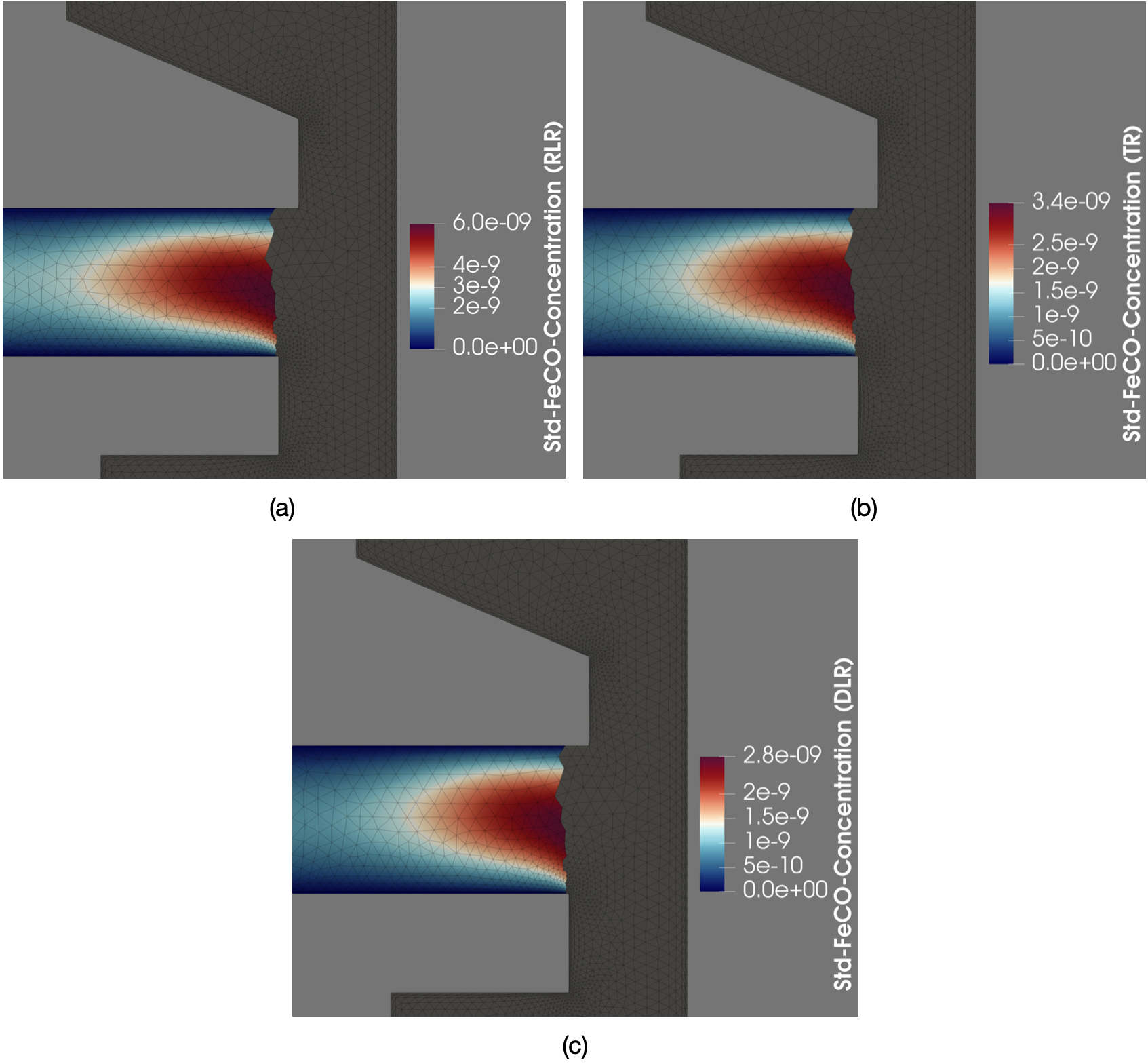}
		    \caption{Spatial distribution over the region near the deposition surface of the standard deviation for Fe(CO) concentrations (P = 10 Torr) in (a) the reaction-limited regime, (b) the transition regime, and (c) the diffusion-limited regime. The color scale on the right indicates darker colors for smaller quantities and lighter colors for larger quantities.
        }\label{S-FeCO}
	\end{figure}

\begin{figure}[!ht]
        \centering
            \includegraphics[width=0.5\linewidth]{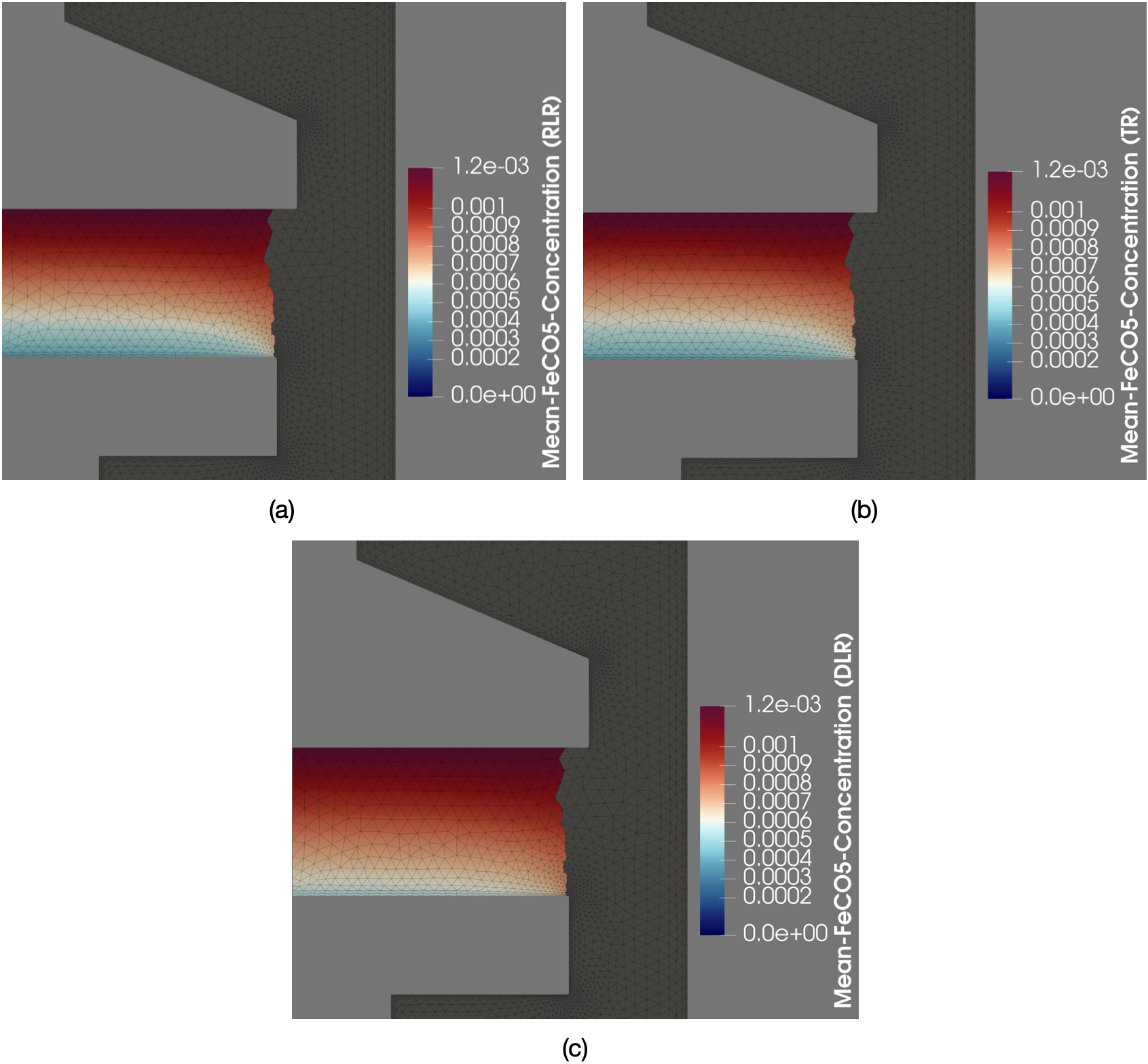}
		    \caption{Spatial distribution over the region near the deposition surface of the mean for Fe(CO)$_5$ concentrations (P = 10 Torr) in (a) the reaction-limited regime, (b) the transition regime, and (c) the diffusion-limited regime. The color scale on the right indicates darker colors for smaller quantities and lighter colors for larger quantities.
        }\label{M-FeCO5}
	\end{figure}

\begin{figure}[!ht]
        \centering
            \includegraphics[width=0.5\linewidth]{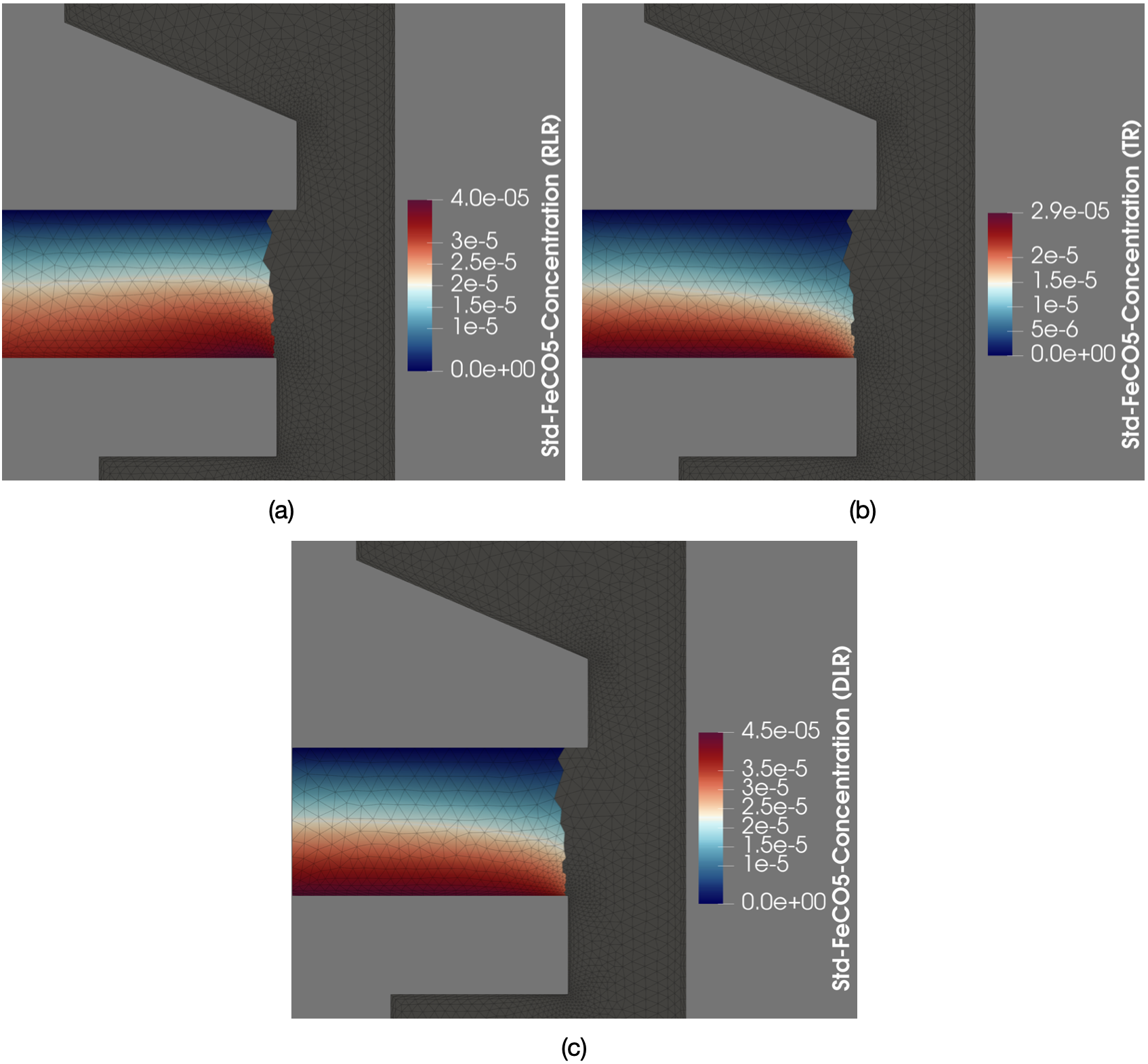}
		    \caption{Spatial distribution over the region near the deposition surface of the standard deviation for Fe(CO)$_5$ concentrations (P = 10 Torr) in (a) the reaction-limited regime, (b) the transition regime, and (c) the diffusion-limited regime. The color scale on the right indicates darker colors for smaller quantities and lighter colors for larger quantities.
        }\label{S-FeCO5}
	\end{figure}

    The first two moments of the model responses, namely the mean and standard deviation, are calculated using the coefficients of the polynomial expansion according to \eqref{eq:m-std}. Their spatial distributions in the study area defined near the deposition surface (cf. \cref{Fig:nodes}) is plotted for Fe(CO) and Fe(CO)$_5$ (cf. Figs. \ref{M-FeCO}, \ref{S-FeCO}, \ref{M-FeCO5} and \ref{S-FeCO5}, respectively), and for each of the regimes identified in \cref{subc:clus}.

    The mean of the Fe(CO) concentrations presented in Fig. \ref{M-FeCO} is lower near the deposition surface in all regimes. Furthermore, the concentration gradient is larger toward the edges of the deposition surface. The values of the standard deviation, shown in Fig. \ref{S-FeCO}, are very small everywhere in the computational domain, signifying a very small deviation of each point in the dataset from the population mean, suggesting lower variability and greater consistency. 

    Further examination of the results for Fe(CO)$_5$ (cf.  Figs.\ref{M-FeCO5} and \ref{S-FeCO5}), reveals even smaller standard deviation overall in the domain, particularly in the vicinity of the deposition surface, suggesting smaller variability.

    The prediction error for the mean and standard deviation for both species is computed in the computational domain. 
    For Fe(CO)$_5$, the relative error for the mean does not exceed $O(-3)$, while the error for the standard deviation is $O(-6)$ throughout the computational domain.
    Similarly, for Fe(CO) the magnitudes for both the mean and the standard deviation are $O(-5)$ on average.

    These results elucidate the effect of uncertainty of the inputs on the distribution of precursor concentrations, in the considered region of interest. The influence of the input parameters (associated with each of the regimes) on the concentrations is revealed by sensitivity analysis in the following section.

\newpage


\newpage
\newpage

 \newpage
 
    \subsection{PCE-based sensitivity analysis}
    The objective of this section is to estimate the effects of the two input parameters on the low-dimensional model responses across the different regimes of the Arrhenius plot, using the Sobol' indices (see \cref{subsubsec:Sobol}).

    Thus, the following results correspond to the Sobol' indices calculated for the two input parameters, $\text{x}_{T_s}$ and $\text{x}_{P}$, with respect to each of the three PC coefficients considered for each regime. In addition, since the sum of the variabilities explained by the PCs for each regime is approximately 100\%, the Sobol' indices are presented using pie charts and should be read as the percentage of variability contributed by an input parameter (and its interactions) to the variability explained by each PC coefficient ($PC_i$).
    
   \subsubsection{Regimes of the Arrhenius plot}
 
    \begin{itemize}
	\item[] \textit{Reaction-limited regime}
    \begin{figure}[ht!]
		\centering
		\includegraphics[width=0.3\linewidth]{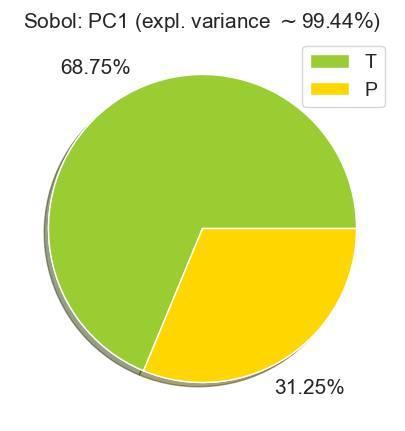}
		\includegraphics[width=0.3\linewidth]{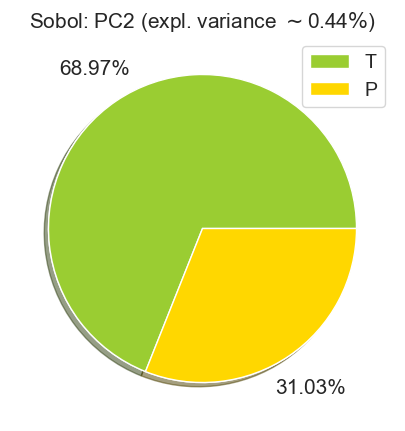}
		\includegraphics[width=0.3\linewidth]{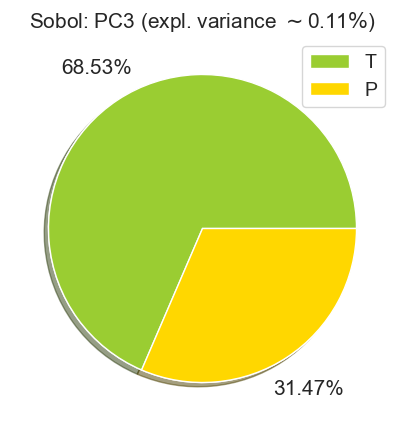}
		\caption{Total Sobol' indices for $T_s$ and $P$ concerning each PC corresponding to the reaction-limited regime (cumulative variance explained $>99.99\ \%$).}\label{fig-Sr1020}
	\end{figure}
 
	In the reaction-limited regime, chemical reaction rates increase with increasing temperature, which explains the relative significance of the substrate temperature $T_s$, as confirmed by the Sobol' indices in \cref{fig-Sr1020}, from which we can infer that the substrate temperature $T_s$ is the most influential parameter in the reaction-limited regime. 

    When we sum the individual contributions, we find that $T_s$ is responsible for $68.74\ \%$ of the total variability, while the remaining $31.25\ \%$ is attributed to the influence of $P$. More precisely, at low(er) temperatures, reactions in the gas phase and on the deposition surface occur at a slower rate than the rate of chemical species diffusion. Therefore, the chemical reaction rate determines the overall deposition rate.  
    \end{itemize}

	\begin{itemize}
	\item[] \textit{Transition regime}

    \begin{figure}[!ht]
		\centering
		\includegraphics[width=0.3\linewidth]{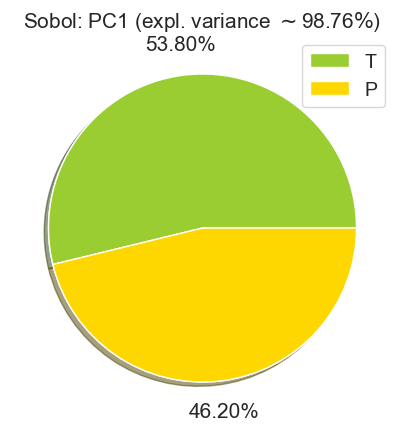}
		\includegraphics[width=0.3\linewidth]{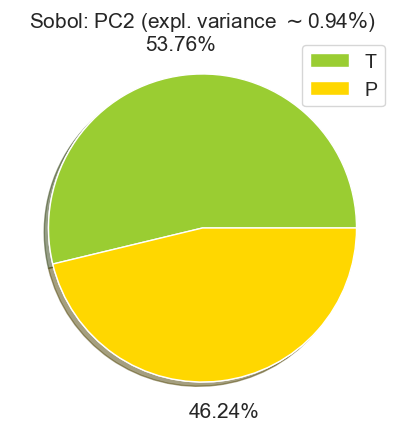}
        \includegraphics[width=0.3\linewidth]{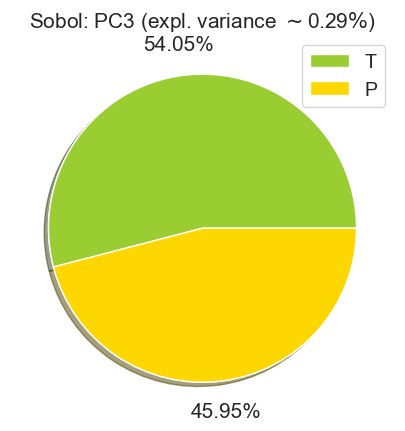}
		\caption{Total Sobol' indices for both $T_s$ and $P$ concerning each PC-mode corresponding to the transition regime (cumulative variance explained $>99.99\ \%$).}\label{fig-St1020}
	\end{figure}
 
    At higher temperature, the reaction rate also increases until it becomes comparable to the species diffusion rate (transition regime). The results of the sensitivity analysis show that the temperature is no longer as important as it was in the reaction-limited regime because pressure gains significance, as it dictates the transport, diffusion, and concentration of the reactive species on and to the substrate.
 
    In \cref{fig-St1020}, the Sobol' indices for this regime are presented. Here, we observe an expected balance of influences produced by the input parameters. More precisely, when we sum up the individual contributions, we obtain the following results: $T_s$ is responsible for $53.80\ \%$ of the total variability, while the remaining $46.19\ \%$ corresponds to the influence of the chamber pressure $P$.
	
	\end{itemize}

	\begin{itemize}
	\item[] \textit{Diffusion-limited regime}

    Further increasing temperatures, typically suggests that the reaction rate is even faster and ultimately the overall reaction rate is determined by the rate of species diffusion on the deposition surface. Nevertheless, in this case, this regime is regulated by gas-phase decomposition reactions that are activated at higher temperatures causing depletion and a decrease in the deposition rate. 
    \begin{figure}[!ht]
		\centering
		\includegraphics[width=0.354\linewidth]{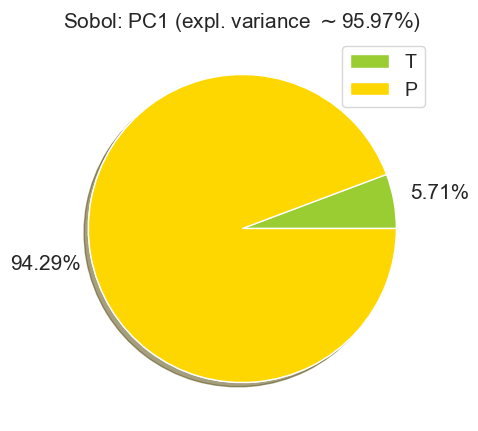}
		\includegraphics[width=0.3\linewidth]{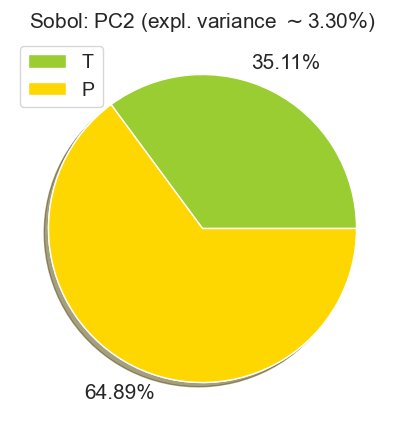}
		\includegraphics[width=0.3\linewidth]{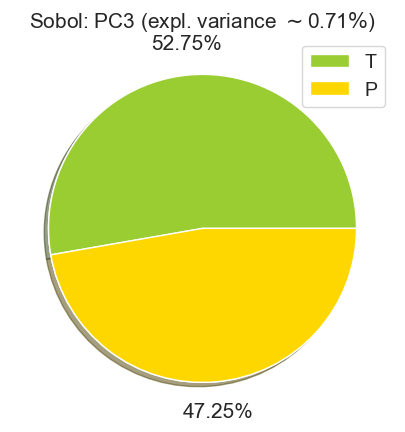}
		\caption{Total Sobol' indices for both $T_s$ and $P$ concerning each PC-mode corresponding to the diffusion-limited regime (cumulative variance explained $> 99.98\ \%$).}\label{fig-Sd1020}
	\end{figure}
 
    Indeed, Sobol' indices (refer to \cref{fig-Sd1020}) reveal that temperature has a small effect, but they also indicate that there exists a greater influence of the chamber pressure. $T_s$ contributes to $7.01\ \%$ of the total variability, while the influence of $P$ corresponds to $92.97\ \%$. 
    
    To interpret these results, we should revisit the Arrhenius plot \cref{Fig:P10} ($P=10$ Torr), where the deposition rate decreases in this region due to two phenomena: (i) The high temperatures cause the precursor to decompose or be depleted in the gas-phase. (ii) There is a species diffusion from the surface back to their gas-phase. Consequently, instead of a typical straight line in the Arrhenius plot, in this case, we have a drop in the deposition rate. 
	\end{itemize}

\section{Conclusions}\label{Sec:Conclusions}
This work presents a collection of data-driven methods able to (i) achieve dimensionality reduction of large data sets resulting from a CFD model of a CVD reactor; (ii) identify clusters of ``outcomes" that correspond to different process regimes, where the relative influence of the various inputs changes, something which is experimentally confirmed by the Arrhenius plot; (iii) create an efficient surrogate model, without sacrificing accuracy; (iv) perform sensitivity analysis to quantify the relative effect of the process inputs on the outcome, in each one of the identified regimes. The outcomes of this analysis aids in the formulation of a hypothesis as to the phenomena taking place past the transition regime: the fact that the temperature is important even in the diffusion-limited regime of the Arrhenius plot, suggests that gas phase reactions are activated at higher temperatures. 

The proposed methods offer reasonable physical insights that align with intuition that results from experimental observations and theoretical knowledge. These insights can be instrumental in decision-making for the design and optimization of the process, bypassing the need for costly and time-consuming experiments. Furthermore, this work demonstrates the potential of computational methods based on data for the design of innovative reactors.


\begin{appendices}
  \crefalias{section}{appsec}
  \section{Appendix: Principal Component Analysis}\label{AppendixA}

Given a dataset $\textbf{Y} (m \times N)$, where we have collected $m$ simulations of the $N$ variables, we start by centering or standardizing the data (to give each variable the same importance throughout the analysis especially when data has different magnitudes) \cite{NAGEL2020, PARENTE13}. This transformation is expressed as follows:
	\begin{align*}
		\tilde{\textbf{Y}} = \left(\textbf{Y} - \bar{\textbf{Y}}\right) \textbf{S}^{-1},
	\end{align*}
	where $\bar{\textbf{Y}}$ is an $m \times N$ matrix containing the mean of each column vector of the simulated output variables and $\mathbf{S}$ is an $m \times N$ matrix containing the corresponding standard deviations ($\mathbf{S} = I_{m \times N}$, when we only center the data).\\
	
	To perform PCA \cite{HOTELLING1933, MACKIEWICZ1993}, the covariance matrix of $\tilde{\textbf{Y}}$ and its SVD are considered:
	\begin{align*}
		\text{cov}(\tilde{\textbf{Y}}) = \nicefrac{\displaystyle \tilde{\textbf{Y}}^{\top} \tilde{\textbf{Y}}}{\displaystyle m-1} = \textbf{V} \mathbf{\Sigma} \textbf{V}^{\top},
	\end{align*}
    where $\textbf{V}$ is an $N \times N$ orthogonal matrix with eigenvectors of $\text{cov}(\tilde{\textbf{Y}})$, whereas the $N \times N$ diagonal matrix $\mathbf{\Sigma} = \text{diag}(\sigma_1, \sigma_2, \cdots, \sigma_N)$ contains the associated eigenvalues arranged in decreasing order, that is, $\sigma_1 \geq \sigma_2 \geq  \cdots \geq \sigma_N$ (see \Cref{Fig:PCA}). More precisely, the columns of matrix $\textbf{V}$ are uncorrelated vectors and correspond to the directions in which the data varies the most, defining a new orthogonal basis. These basis vectors are called \textit{principal components} (PCs). Similarly, we can interpret the corresponding eigenvalues as variances of the data in such directions.
    
    Subsequently, we projected the standardized dataset $\tilde{\textbf{Y}}$ on the new basis. The obtained values define the \textit{principal component scores} arranged in an $m \times N$ matrix $\textbf{Z} = \tilde{\textbf{Y}} \textbf{V}$. In addition, since the variance captured by each PC decreases, it is sufficient to retain only the first $n$ dominant PCs that describe the highest cumulative variability. Thus, from $\tilde{\textbf{Y}}$ and $\textbf{V}$, we can compute their truncated versions $\tilde{\textbf{Y}}_n$ ($m \times n$) and $\textbf{V}_n$ ($N \times n$) such that
	\begin{align}\label{Approxn}
		\textbf{Z} \approx \tilde{\textbf{Z}}_n = \tilde{\textbf{Y}}_n \textbf{V}_n.
	\end{align}
	
	A criterion for finding the smallest $n \leq N$, that suffices for the accurate data representation, is to study the total variance ratio such that it is larger than a fixed threshold $\delta > 0$, i.e., $\nicefrac{\sum_{i = 1}^{n} \sigma_i}{\sum_{i = 1}^{N} \sigma_i} > \delta$. We retain the PCs required to explain at least $99\%$ of the total variance of the data.
    Furthermore, using equation \eqref{Approxn}, the original dataset can be reconstructed as follows:
	\begin{align}\label{eq:recons}
		\tilde{\textbf{Y}} \approx \tilde{\textbf{Z}}_n \textbf{V}_n^{\top},
	\end{align}
	where $\textbf{V}_n^{\top}$ ($m \times n$) denotes the transpose of $\textbf{V}_n$.\\ 

    Additionally, to assess the accuracy of PCA, we use equation \eqref{eq:recons} and compute the  \textit{reconstruction error} by means of the \textit{mean squared error}, $\text{MSE} = \mathbb{E} \left[\big(\tilde{\textbf{Y}} - \tilde{\textbf{Z}}_n \textbf{V}_n^{\top}\big) \big(\tilde{\textbf{Y}} - \tilde{\textbf{Z}}_n \textbf{V}_n^{\top}\big)^{\top}\right]$.

\end{appendices}

\section*{CRediT authorship contribution statement}
\textbf{Geremy Loachamín Suntaxi}: Writing - review $\&$ editing, Writing - original draft, Visualization, Validation, Methodology, Formal analysis, Data curation, Conceptualization. 
\textbf{Paris Papavasileiou}: Writing - review $\&$ editing, Writing - original draft, CFD modeling and data collection. 
\textbf{Eleni D. Koronaki}: Writing - review $\&$ editing, Writing - original draft, Validation, Supervision, Project administration, Investigation, Funding acquisition, Formal analysis, Conceptualization. 
\textbf{Dimitrios G. Giovanis}: Writing - review $\&$ editing, Validation, Methodology, Conceptualization. 
\textbf{Ioannis G. Aviziotis}: Writing - review $\&$ editing, Validation, Formal analysis.
\textbf{Georgios P. Gakis}: Writing - review $\&$ editing, Writing - original draft, Formal analysis.
\textbf{Gabriele Pozzetti}: Writing - review $\&$ editing, Resources.
\textbf{Martin Kathrein}: Writing - review $\&$ editing, Resources, Project administration.
\textbf{Christoph Czettl}: Writing - review $\&$ editing, Resources.
\textbf{Andreas G. Boudouvis}: Review $\&$ editing, Supervision.
\textbf{Stéphane P.A. Bordas}: Writing - review $\&$ editing, Supervision, Project administration, Funding acquisition.

\section*{Declaration of competing interest}
The authors declare that they have no known competing financial interests or personal relationships that could have appeared to influence the work reported in this paper.

\section*{Data availability}
Data will be made available on request.

\section*{Acknowledgements}
This research was funded by the Luxembourg National Research Fund (FNR), grant reference 16758846. For the purpose of open access, the authors have applied a Creative Commons Attribution 4.0 International (CC BY 4.0) license to any Author Accepted Manuscript version arising from this submission.

\end{document}